# Evaluating the Sensitivity of Mortality Attributable to Pollution to Modeling Choices: A Case Study for Colorado

Priyanka N. deSouza[1,2,3,*], Susan Anenberg[4], Neal Fann[5], Lisa M. McKenzie[6], Elizabeth Chan[5], Ananya Roy[7], Jose L. Jimenez[8,9], William Raich[10], Henry Roman[10], Patrick L. Kinney[11]

1: Department of Urban and Regional Planning, University of Colorado Denver, Denver, CO, USA
2: CU Population Center, University of Colorado Boulder, CO, USA
3: Senseable City Lab, Massachusetts Institute of Technology, USA
4: Miliken Insittute School of Public Health, George Washington University, Washington D.C., USA
5: U.S. Environmental Protection Agency
6: Department of Environmental and Occupational Health, Colorado School of Public Health, University of Colorado Anschutz, Aurora, CO, USA
7: Environmental Defense Fund, NY, USA
8: Cooperative Institute for Research in Environmental Sciences, University of Colorado Boulder, Boulder, CO, USA
9: Department of Chemistry, University of Colorado Boulder, Boulder, CO, USA
10: Industrial Economics, Boston, MA, USA
11: Boston University School of Public Health, MA, USA

*: Corresponding author (priyanka.desouza@ucdenver.edu)

# Abstract

We evaluated the sensitivity of estimated $PM_{2.5}$ and $NO_2$ health impacts to varying key input parameters and assumptions including: 1) the spatial scale at which impacts are estimated, 2) using either a single concentration-response function (CRF) or using racial/ethnic group specific CRFs from the same epidemiologic study, 3) assigning exposure to residents based on home, instead of home *and* work locations. This analysis was carried out for the state of Colorado. We found that the spatial scale of the analysis influences the magnitude of $NO_2$, but not $PM_{2.5}$, attributable deaths. Using county-level predictions instead of 1 $km^2$ predictions of $NO_2$ resulted in a lower estimate of mortality attributable to $NO_2$ by ~ 50% for all of Colorado for each year between 2000-2020. Using an all-population CRF instead of racial/ethnic group specific CRFs results in a higher estimate of annual mortality attributable to $PM_{2.5}$ by a factor 1.3 for the white population and a lower estimate of mortality attributable to $PM_{2.5}$ by factors of 0.4 and 0.8 for Black and Hispanic residents, respectively. Using racial/ethnic group specific CRFs did not result in a different estimation of $NO_2$ attributable mortality for white residents, but led to lower estimates of mortality by a factor of ~ 0.5 for Black residents, and by a factor of 2.9 for to Hispanic residents. Using $NO_2$ based on home instead of home *and* workplace locations results in a smaller estimate of annual mortality attributable to $NO_2$ for all of Colorado by ~0.980 each year and 0.997 for $PM_{2.5}$.

# 1 Introduction

Research has consistently highlighted exposure to ambient air pollution as an important contributor to death and disability (Boing et al., 2022; deSouza et al., 2020; deSouza et al., 2022a, 2022b). Overall, air quality in the US has improved dramatically since the adoption of the 1970 Clean Air Act (CAA) (Krupnick and Morgenstern, 2002). However, racial and ethnic disparities in exposure to pollution have persisted (Colmer et al., 2020). Lower income, minority and marginalized populations experience higher air pollution exposure levels and associated health impacts (deSouza et al., 2023, 2020; deSouza et al., 2022; Josey et al., 2023).

Decision makers often evaluate the effectiveness of air pollution policies on the basis of health impact assessments (HIA) of air pollution reductions, i.e. evaluating the number of adverse health outcomes avoided by improving air quality (Fann et al., 2013; Hubbell et al., 2009). Most nation-wide analyses used by U.S EPA to quantify mortality risks associated with changes in air pollutants are at the county-scale (Fann et al., 2018), and employ a single concentration-response function (CRF) between exposure to a pollutant and the health outcome. Associations between fine particulate matter [aerodynamic diameter ≤ 2.5 μm; $PM_{2.5}$], for example are often adopted from the American Cancer Study (ACS) (Krewski et al., 2009), the Harvard Six Cities analyses (Lepeule et al., 2012), more recent work using the Medicare cohort (Wu et al., 2020), and studies conducted using data from the National Health Interview Surveys (Pope et al., 2019). However, as these studies evaluated the impacts of pollution in populations comprised of people of socioeconomic status (SES) above the national average, and were predominantly White in urban locations, using the CRFs derived from these studies likely underestimates the health impacts of pollution in lower income, minority and marginalized communities. Research has demonstrated that using race/ethnicity-specific CRFs results in significantly larger benefits of $PM_{2.5}$, reduction policies for Black Americans (Spiller et al., 2021).

Other research has shown that the spatial scale of the HIA of pollution can also impact the results. A study conducted in India found that results of the impact of $PM_{2.5}$ on premature mortality obtained from using a uniform nation-wide baseline mortality rate across India (as considered in the Global Burden of Disease), instead of state-specific baseline mortality rates was lower by ~ 15% (Chowdhury and Dey, 2016). Castillo et al., (2021) found that using more spatially disaggregated baseline disease rates at the neighborhood instead of ward-level in Washington D.C. yielded more variation in $PM_{2.5}$ attributable mortality rates, which more effectively highlighted environmental injustices in D.C. Castillo et al., (2021) also demonstrated that using baseline disease rates (BDR) from local administrative records, instead of estimates of BDR from the CDC 500 Cities project, also yielded different patterns in the variation of the health impacts of $PM_{2.5}$ for D.C. Specifically, HIA results from the administrative health records yielded larger racial and ethnic disparities than those from the CDC 500 cities dataset.

Research conducted in the San Francisco Bay area of California found that using census block-group (CBG) level baseline mortality rates, instead of county-level disease rates resulted in a 15% larger attributable mortality rates for the pollutants $PM_{2.5}$ and nitrogen dioxide ($NO_2$) (Southerland et al., 2021).

Southerland et al., (2021) also found that using granular estimates of pollution at a 100 m x 100 m resolution from a mobile monitoring campaign, instead of widely-used 1 km x 1 km satellite-derived estimates of $PM_{2.5}$ and $NO_2$ yielded significantly larger spatial heterogeneity in attributable mortality rates that revealed more spatial heterogeneity in pollutant-attributable health risks in the San Fancisco Bay area. This issue may be especially important when assessing health impacts in low-income, minority communities, where fine scale spatial gradients often exist in both demographics and air pollution concentrations. While several studies have explored the influence of spatial resolution of pollutant concentrations on estimated pollution-attributable health impacts (Fenech et al., 2018; Jiang and Yoo, 2018; Li et al., 2016; Mohegh et al., 2020; Parvez and Wagstrom, 2020; Punger and West, 2013), there is a paucity of research that has evaluated the sensitivity of the HIA to the spatial resolution of the health data used.

In the present study, we evaluated the health impact of two pollutants: $PM_{2.5}$ and $NO_2$ on annual all-cause mortality in the state of Colorado between 2000-2020. We considered the outcome: all-cause mortality, as it has been determined to be causally associated with both pollutants (Josey et al., 2022; Wei et al., 2021). $PM_{2.5}$ is a regional pollutant and doesn't often exhibit large local variation within cities (deSouza et al., 2020). $NO_2$, however, is a traffic-related pollutant and can vary significantly over small areas (Apte et al., 2017). $PM_{2.5}$ is the pollutant thought to be associated with largest health burden (and economic cost) of all air pollutants (Tschofen et al., 2019). Although $NO_2$ has been linked with adverse health outcomes, it is often not quantified in pollution attributable-studies, potentially because coarsely resolved concentration estimates are often unable to capture highly spatially variable patterns in $NO_2$ (Southerland et al., 2021). However, the ability to quantify the health impacts of $NO_2$ has become more robust with recent advances in $NO_2$ exposure assessments and a better understanding of the health effects of $NO_2$, including meta-analyses and published recommendations from a committee of scientists on evaluating and interpreting $NO_2$ as a marker of the mixture of traffic air pollution (Atkinson et al., 2018; Huangfu and Atkinson, 2020; Khreis et al., 2017; Thurston et al., 2020).

We compared state-wide administrative mortality rates for the state of Colorado at the county-level with that derived from modeled estimates from the CDC Wonder database. We then evaluate the sensitivity of HIA results in Colorado from 2000-2020 to key modeling choices, i.e., (1) the spatial scale of the baseline health outcome dataset and pollution exposure assessment (block, block group, census tract, and county levels), (2) the choice of CRF (a uniform CRF versus race and ethnic specific CRFs), and (3) assigning exposures to residents based on home *and* work-locations, instead of just home-based exposures. We also assess racial and ethnic disparities in health impacts of pollution to under sensitivity analyses (1) and (2). We did not do so for (3) because we do not have racial/ethnic group specific exposures considering home and work-locations.

Colorado has a population of ~5.7 million (~2021) up from ~ 4.3 million in 2000, with a median age of 36.9 years, a median household income of USD 75,231. White (Non-Hispanic) residents are the majority of the population ~ 67.5%, followed by White (Hispanic) ~ 14%. Between 2000 and 2021 the share of the Hispanic population increased by 5% points. Our results can inform the choice of spatial scale in future HIAs as well as in assessing the distributional benefits of future policies enacted as part of the Justice40 initiative that has concerns about environmental justice at its core (Siddiqi et al., 2022).

# 2 Methods and Materials

## 2.1 Pollution Concentrations

### 2.1.1 $PM_{2.5}$ concentrations

We use annual mean $PM_{2.5}$ concentration estimates (units: µg/m$^3$) from a North American satellite-derived data set (V5.GL.03) with a spatial resolution of 0.01° × 0.01° (~1 km$^2$) for each year beginning in 2000 through 2020 (Hammer et al., 2020). This data set relates the combined aerosol optical depth (AOD) from multiple satellite retrievals to surface $PM_{2.5}$ concentrations using the spatiotemporally varying geophysical relationship between AOD and $PM_{2.5}$ simulated by the GEOS-Chem chemical transport model. These geophysical values are calibrated to ground-based monitors using a geographically weighted regression (Hammer et al., 2020) (**Figure 1A**).

### 2.1.2 $NO_2$ concentrations

We used annual $NO_2$ concentrations (units: ppbv) at a 0.0083° x 0.0083° (~1 km$^2$) scale for the years 2000, 2005 and each year from 2010 through 2020. These previously published estimates were created by adjusting an existing global $NO_2$ dataset representing average concentrations at a 100 m resolution for the years 2010-2012 (Larkin et al., 2017) to correct for high bias in rural areas, and then further extended to other years using observations from the Ozone Monitoring Instrument (OMI) aboard NASA's Aura satellite (Anenberg et al., 2022) (**Figure 1D**).

We estimated census block, block-group, tract and county-level exposures (using the 2010 census tract boundaries) to annual-average $PM_{2.5}$ and $NO_2$ concentrations using a spatially weighted mean of the grid cells within a census tract using the exact_extract package (Baston et al., 2022) in R 4.2.3. The average area of a census block in Colorado is 1.35 km$^2$ (min: 0.00, 25th percentile: 0.01 km$^2$, median: 0.02 km$^2$, 75th percentile: 0.16 km$^2$, max: 1099.2 km$^2$), while that of a block-group is 76.7 km$^2$ (min: 0.07 km$^2$, 25th percentile: 0.52 km$^2$, median: 0.99 km$^2$, 75th percentile: 4.2 km$^2$, max: 9034.4 km$^2$), and that of a census-tract is 216.8 km$^2$ (min: 0.11 km$^2$, 25th percentile: 1.96 km$^2$, median: 216.8 km$^2$, 75th percentile: 23.01 km$^2$, max: 11,988.5 km$^2$). We note that the spatial scale of exposure assessment is generally more coarse than that of the spatial unit under consideration. However, the exposure datasets that we rely on are the most spatially granular, publicly available for the entire state.

We also display 1 km x 1 km $PM_{2.5}$ and $NO_2$ concentrations for all of Colorado for the years 2015 and 2020 in **Figure S1**.

### 2.1.3 Assigning exposure based on home *and* work census tracts

We evaluated the sensitivity of assigning exposure according to residential, instead of assigning exposure according to the mobility patterns of residents. Specifically, we used the LODES (version 7.5) product (https://lehd.ces.census.gov/data/lodes/LODES7/, Last accessed December 8, 2022) from the US Census Bureau for every year between 2002 and 2019 to evaluate exposures based on home and workplace locations. The LODES data derive from administrative records (e.g., state employment insurance reporting and federal worker earnings records) of home and work addresses of individual workers (aged 14 and over) and are aggregated to the home and work census blocks for a representative sample of workers for every state. The data cover ~ 95% of jobs in the United States. We specifically used the LODES Origin-Destination (OD) dataset which provides information on the number of individuals commuting between home and primary work census blocks. We aggregated the LODES OD data to the census-tract level to be consistent with the resolution of the pollution data we use. In this analysis we only considered primary jobs, alone, and not secondary or tertiary jobs. We also restricted our analysis to workers living *and* working in the state of Colorado.

We calculated an annual average population-weighted exposure (H) using **Equation 1**:

$$\overline{PM_{2.5,H}} = \frac{\sum_h PM_{2.5,h} p_h}{\sum_h p_h} \quad (1)$$

where $PM_{2.5,h}$ denotes the $PM_{2.5}$ concentrations for residential census tract *h* (H); $p_h$ signifies the number of workers residing in home census tract *h*. When evaluating population-weighted exposure using workplaces (W), $PM_{2.5,w}$ concentrations for work census tract *w* were used (W) in **Equation 1.** $p_h$ is replaced by $p_w$ corresponding to the number of workers working in census tract *w*.

For the home-workplace exposure metric (HW), we assumed that individuals were in their workplace census tract for 1,801 hours per year (based on an 8 hour work day, 5 days a week, 45 weeks per year) out of a total of 8760 hours per year (20.6%), and we used their home census tract for the remaining time (deSouza et al., 2023). We thus evaluate their HW exposure as 79.4% of H + 20.6% of W. We calculate population-weighted HW exposure using **Equation 2**:

$$\overline{PM_{2.5,HW}} = 0.794 \times \overline{PM_{2.5,H}} + 0.206 \times \overline{PM_{2.5,W}} \quad (2)$$

$PM_{2.5,HW}$ is displayed in **Figure 1**.

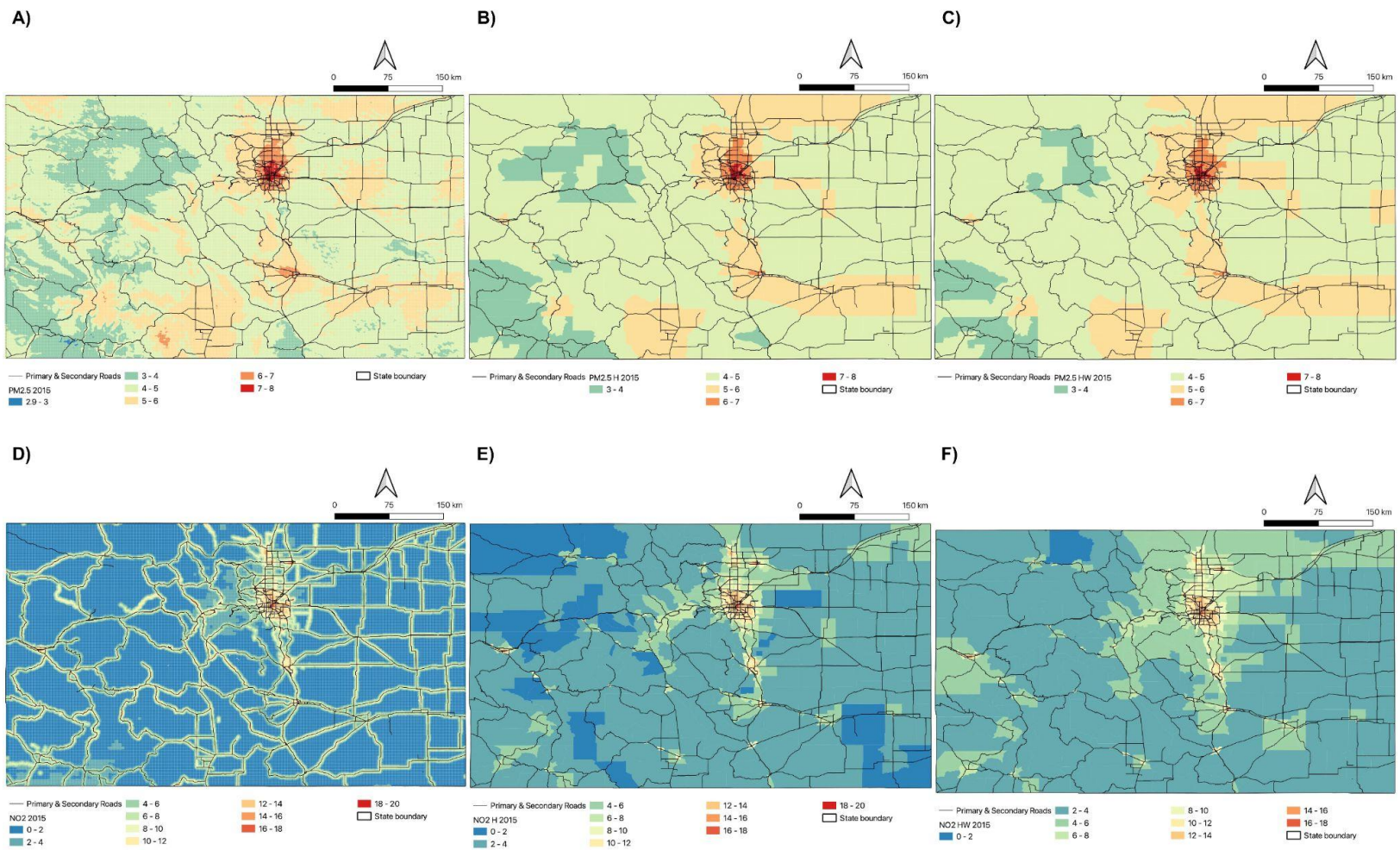

*Figure 1*: Annual-averaged A) 1 km x 1 km PM$_{2.5}$ (μg/m$^3$), B) Census-tract level PM$_{2.5}$ (μg/m$^3$), C) Census-tract level population-weighted PM$_{2.5}$ (μg/m3) *considering home and work-place exposures*, D) 1 km x 1 km NO$_2$ *levels (ppbv)*, E) Census-tract level NO$_2$ *levels (ppbv)*, F) Census-tract level population-weighted NO$_2$ *levels (ppbv) considering home and work-place exposures over Colorado for the year 2015.*

## 2.2 Baseline Disease Rates and Counts

### 2.2.1 Administrative Baseline Mortality Counts (BMC)

We obtained annual baseline all-cause mortality counts (BMC) at the census block-level for the years 2000-2020 from the Colorado Department of Public Health and Environment (CDPHE) by race and ethnicity (White non-Hispanic, Hispanic all races, Black non-Hispanic, Asian or Pacific Islander non-Hispanic, American Indian non-Hispanic, Other non-Hispanic, Unknown). Note that every year ~4% of all records had an unknown location of death. These were excluded from the analysis.

We obtained total county-level mortality data between 2000-2020 for all-cause mortality from CDC Wonder (CDC, 2023). We also obtained county-level mortality data by the same racial and ethnic categories as the administrative data for Colorado. We compared the CDC Wonder data with the administrative data we obtained from CDPHE. We also compared the two datasets for counties corresponding to different Colorado Enviroscreen percentiles (an environmental justice mapping tool that indicates more vulnerability).

## 2.3 Population Data

We obtained total population estimates from the Gridded Population of the World (GPW), Version 4, by the Center for International Earth Science Information Network (CIESIN) available at a ~ 1 km$^2$ spatial resolution from the Socioeconomic Data and Applications Center (SEDAC) for the years 2000, 2005, 2010, 2015 and 2020. We linearly interpolated population estimates for other years in the range 2000-2020 using this dataset. Just as for the pollution data, we evaluated census block-, block-group, tract, and county-level population estimates in Colorado. For racial/ethnic group-specific population data at the block-group level we used data from the 2020 decennial census, and at the block-level from the 2010 decennial census (latest data available).

We display overall baseline mortality rates (All-cause mortality/total population from GPW v4; BMR) at different spatial scales in **Figure 2**. Note we do not display BMR at the block-level because the population data is more coarse than the block-spatial scale for Colorado. **Figure S2** displays the total BMC between 2000-2020 at the block-level for Colorado.

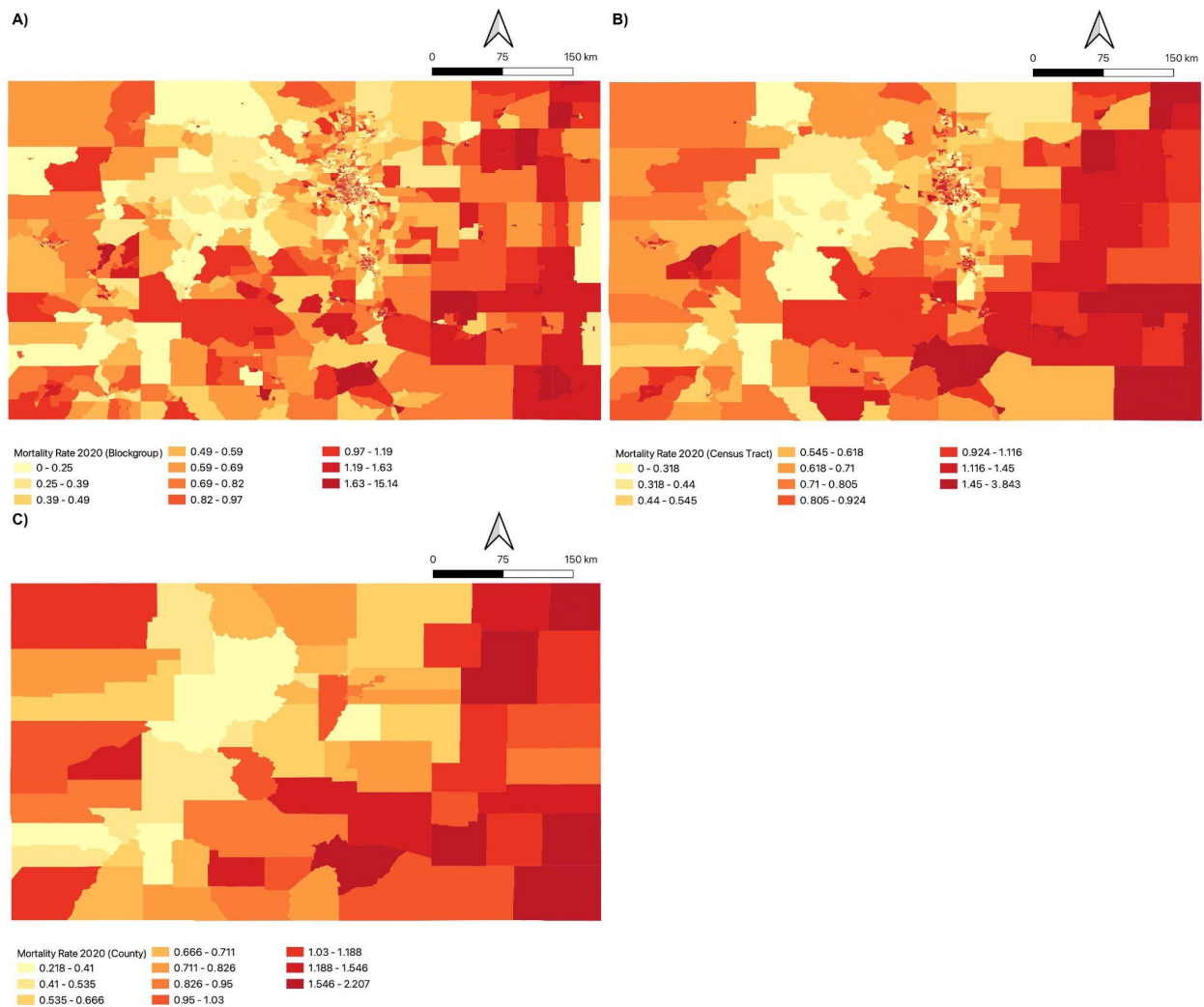

*Figure 2*: All-cause Mortality Rate (%) for the total population in the year 2020 at the A) block group, B) census tract, and C) county scales, classified into deciles.

## 2.4 Evaluating spatio-temporal variation in BMR derived from administrative data

We used a multilevel model to partition variation in BMR (obtained by dividing BMC by census-block level population data, discussed in *section 2.3*) across different spatial and temporal scales. Specifically, we analyzed the variation in BMR in Colorado with multilevel linear models, including random effects for year, county, census-tract, block group, and block. We report the crude variation in BMR at each spatial and temporal level. The proportion of variance attributed to each level, z, was computed as follows: ($var_z$ / ($var_{year}$ + $var_{county}$ + $var_{census\ tract}$ + $var_{block\ group}$ + $var_{block}$) × 100).

## 2.5 Evaluating pollution-attributable mortality

We apply widely-used epidemiologically-derived CRFs, applied uniformly across racial/ethnic groups to estimate mortality attributable to annual-average $PM_{2.5}$ and $NO_2$. We use log-linear relationships between $PM_{2.5}$ concentrations and all-cause mortality relative risk (RR): 1.06 (95% CI: 1.04, 1.08) per 10 µg/m³ increase in $PM_{2.5}$ from Turner et al., (2016) consistent with previous studies (Castillo et al., 2021), while that between $NO_2$ and all-cause mortality was 1.02 (95% CI: 1.01, 1.04) per 10 µg/m³ increase in $NO_2$ (Huangfu and Atkinson, 2020). A secondary systematic review reported that this association was 1.04 (95% CI: 1.01, 1.06) (HEI, 2022). We choose the more conservative RR for this analysis. Note for $NO_2$, our pollution dataset had units of ppbv, we used a conversion factor of 1 ppbv = 1.88 µg/m³. We note that this conversion factor was developed at 298 K at sea-level. Given the lower ambient pressure in Colorado, this conversion factor has some error, and will vary across the state. However, as we are interested in evaluating the sensitivity of our results to different modeling choices, the conversion factor will not impact the comparison of our estimates.

In sensitivity analyses, we rely on racial/ethnic specific CRFs from two studies that report these coefficients (Di et al., 2017; Eum et al., 2022; listed in **Table 1**) to evaluate the sensitivity of HIA estimates to using subpopulation specific CRF, instead of an overall CRF. Both studies were conducted on the nation-wide Medicare population, adults over 65 years of age in the United States, a population which is different from the overall population in Colorado. Moreover, the Eum et al., (2022) study reports racial/ethnic specific CRFs for non-accidental deaths and not all-cause deaths, which our study considers. However, due to the dearth of information on subgroup specific CRFs, we use the CRFs reported in these studies. The only other study that considers racial/ethnic specific CRFs for evaluating the mortality attributable to $PM_{2.5}$ also used the CRFs reported in Di et al., (2017) (Spiller et al., 2021). Note that the overall CRF reported in these studies is different from the overall CRF used in the main analysis. For internal consistency, when comparing mortality attributable to $PM_{2.5}$ and $NO_2$ using subpopulation-specific CRFs instead of an overall-estimate, we used the overall estimates reported in **Table 1**.

*Table 1*: Concentration response functions associated with 10 µg/m³ of annual $PM_{2.5}$ from Di et al., (2017) and 10 ppbv increase in annual $NO_2$ levels from Eum et al., (2022).

| **Racial/Ethnic Group** | **Hazard Ratios (95% CI)** | |
|---|---|---|
| | **$PM_{2.5}$** | **$NO_2$** |
| **All** | 1.073 (1.071, 1.075) | 1.06 (1.06, 1.07) |
| **White** | 1.063 (1.060, 1.065) | 1.08 (1.08, 1.09) |
| **Black** | 1.208 (1.199, 1.217) | 1.13 (1.13, 1.14) |
| **Hispanic** | 1.096 (1.075, 1.117) | 1.02 (1.01, 1.03) |
| **Asian** | 1.116 (1.100, 1.133) | 1.05 (1.01, 1.03) |

| Native American | 1.100 (1.060, 1.140) | - |

We evaluate pollution attributable mortality by conducting and comparing separate analyses at the 1) block, 2) block group, 3) census-tract and 4) county-level. For each analysis we aggregate exposures and BMCs to the corresponding spatial scale. For each spatial unit of analysis in Colorado, we estimate the annual excess cases of mortality that are attributable to $PM_{2.5}$ using Equation 1:

$$Mort = (1 - e^{-\beta x}) \times BMC \quad (1)$$

Where β is a mortality specific concentration-response factor from the relative risks (RR) derived from epidemiologic work, $x$ represents $NO_2$ or $PM_{2.5}$ concentrations at the spatial unit of analysis, BMC represents the baseline mortality count (baseline mortality rate (BMR) × population) for the spatial unit of analysis. We use log-linear relationships between the concentration and RR, consistent with previous studies. We display annual mean $NO_2$ and $PM_{2.5}$-attributable excess mortality for the years 2000, 2005, 2010-2020 and 2000-2020, using annual BMC and exposure data, at the block, block-group, census tract, and county-level in a way that is meaningful to inform policy. In further sensitivity analyses, we also conduct the HIA at the spatial scale of the pollution data (1 km x 1 km grid cells). We assign the BMR for each 1 km x 1 km grid cell as the spatially weighted mean BMR from a) block, b) block-group, c) census-tract, and d) county-level BMR estimates using the exact_extractr package in R (Baston et al., 2022).

We sum up the annual total estimated mortality attributable to $PM_{2.5}$ and $NO_2$ over the entire state of Colorado using pollution and health data at the different spatial scales described, and compare the total estimates obtained. We also compare the sensitivity of the total % mortality attributable to pollution of total all-cause mortality to the spatial scale of the health data. We repeat these calculations using race/ethnicity specific mortality data and compare race specific estimates of mortality attributable to pollution using different spatial scales of analyses. We also test the sensitivity of overall, and race specific mortality attributable to using racial and ethnic specific CRFs.

Finally, instead of using pollution data based on residential census tract, we re-ran our analysis using pollution data based on residential and work-place census tract information from the LODES dataset described previously. We compare the census-tract specific mortality attributable to pollution, as well as the total mortality attributable to pollution over the entire domain of Colorado. As the LODES OD dataset does not have race specific information, we only perform this comparison for total mortality.

# 3 Results

**Figure 3** displays annual mean BMC and BMR, as well as $PM_{2.5}$ and $NO_2$ concentrations between 2000-2020 for Colorado (tabulated in **Table S1**). BMR begins to display a sustained increase from 2010 from 0.60% to 0.77% in 2020. The steep increase in 2020 in BMC and BMR is a result of COVID. Although the CDC and administrative data track well ($R^2 \sim 1$), administrative BMCs and BMRs are lower due to the number of records for which no location was available (**Figure S3A**). **Figure S3B** also shows good agreement between county-level administrative and CDC race-specific BMC for all racial and ethnic groups considered in this study ($R^2 \sim 1$). **Table S2** displays BMC and BMR by race from both data sources. It appears that white, non-Hispanic residents have the highest crude (not age adjusted) BMR compared with other racial/ethnic groups in Colorado. We observed higher administrative county-level BMRs across all races for counties that correspond to a higher Colorado Enviroscreen percentile (**Figure S4**). Block-level BMC between 2000 - 2020 is displayed in **Figure S2**. **Figure S5** displays the spatial distribution of Colorado's Enviroscreen score and the fraction of non-white residents.

Colorado follows national trends of decreases in $PM_{2.5}$ (from 5.5 µg/m³ in 2000 to 4.2 µg/m³ in 2019; population-weighted concentrations were 7.1 µg/m³ in 2000 and 5.7 µg/m³ in 2019) and $NO_2$ (from 4.2 ppbv in 2000 to 2.9 ppbv in 2019; population-weighted concentrations were 17.2 ppbv in 2000 and 8.9 ppbv in 2019 concentrations over time). The increase in $PM_{2.5}$ in 2020 is likely a consequence of wildfires the region experienced (deSouza et al., 2023). We used the 1 km × 1 km pollution and population data aggregated over Colorado to calculate our main exposures of interest (**Figure 3**).

We also use LODES-census tract level population and mean census-tract level $PM_{2.5}$ concentrations to report mean $PM_{2.5,H}$ and $PM_{2.5,HW}$ exposures aggregated over all of Colorado. Despite LODES being a subset of the overall population in each census-tract, the main population-weighted $PM_{2.5}$ and $NO_2$ concentrations are not substantially different from the LODES population-weighted $PM_{2.5,H}$. Mean $PM_{2.5,HW}$ is slightly higher than $PM_{2.5,H}$ across years because $PM_{2.5,W}$ is generally higher than $PM_{2.5,H}$.

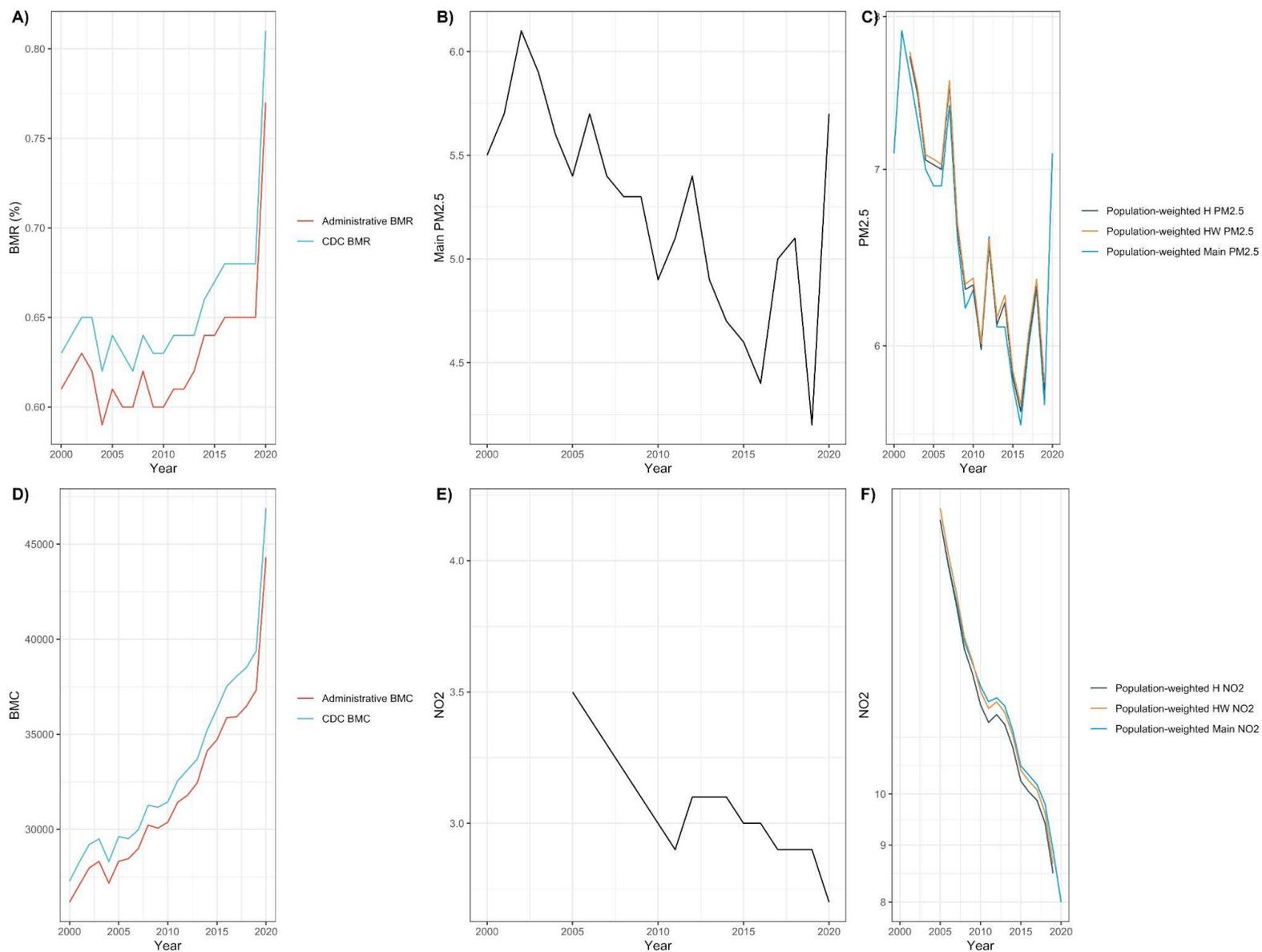

*Figure 3*: Colorado Annual-averaged A) BMR calculated from Administrative and CDC Wonder Data, B) $PM_{2.5}$ (µg/m$^3$) derived using concentration data at a 1 km x 1 km resolution, C) Population-weighted Main $PM_{2.5}$ (µg/m$^3$) *using concentration and population data at a 1 km x 1 km resolution over Colorado and Population-weighted $PM_{2.5}$, exposure levels using census-tract concentration levels for residential ($PM_{2.5,H}$) and census-tract level population data for residential and work ($PM_{2.5,HW}$)* from the LODES dataset *, D) BMC calculated from Administrative and CDC Wonder Data, E) $NO_2$ (ppbv) and F) Population-weighted $NO_2$ (ppbv) levels using concentrations and population data at a 1 km x 1 km resolution over Colorado and population-weighted $NO_{2,H}$ and $NO_{2,HW}$ levels using census-tract concentration levels and LODES population data.*

There is large variation in mortality rates across Colorado (**Figure 2**). The degree of heterogeneity in mortality rates depend on the geographic scale of analysis. For example, in 2020, the blockgroup BMR ranged between 0.25%-15.14%. Less variation in the BMR range (0.22% - 2.21%) was observed at the county-level (**Figure 2**). Our multilevel models indicate that, proportionally, spatial factors (68%) account for more variance in BMR than temporal factors (32%). The variance captured by spatial factors increases with spatial resolution with 47.4%, 11.95%, 6.35%, and 1.78% of proportional variance in BMR accounted for at the block, block group, census tract and county levels, respectively (**Table 2**).

*Table 2: Variance estimates (SEs) in BMR, and the proportion of variance attributable to year, county, census tract, block group, and block-levels*

|  | Year | County | Census Tract | Block group | Block |
|---|---|---|---|---|---|
| **Variance (SE)** | 0.296 | 0.016 | 0.058 | 0.109 | 0.432 |
| **% Variance Explained at the different levels considered** | 32.5% | 1.78% | 6.35% | 11.95% | 47.42% |

We found that the total mortality attributable to $PM_{2.5}$ summed over all of Colorado, when using $PM_{2.5}$ concentrations at their native 1 km × 1 km resolution, were relatively insensitive to the spatial-resolution of the BMR. Mean mortality attributable to $PM_{2.5}$ when using block-level BMR data was higher by ~ 10 % than when using county-level BMR estimates, across years. For example, in 2000, using $PM_{2.5}$ at its native resolution paired with county-level BMR data led to an estimated 1080 (95% CI: 732, 1417) deaths attributable to $PM_{2.5}$, while paired with block-level BMR data led to an estimated 1183 (95% CI: 802, 1551) deaths attributable to $PM_{2.5}$. We also observed that the spatial scale of the $PM_{2.5}$ data chosen also did not have a large impact on the total HIA estimate, with using county-level $PM_{2.5}$, instead of 1 km × 1 km $PM_{2.5}$ estimates resulting in ~ 10% decrease in attributable mortality across years. For example, in 2000, using county-level $PM_{2.5}$ and county-level BMR data led to a total of 953 (95% CI: 646, 1252) deaths attributable to $PM_{2.5}$ (**Figure 4**, **Table S2**).

Mortality impacts attributable to $NO_2$ were more sensitive to spatial resolution of the exposure data than was the case for $PM_{2.5}$. Using less spatially resolved BMR (county instead of block-level) results in a ~10% decrease in the estimated attributable mortality. For example, in 2000, using $NO_2$ at it's native resolution paired with county-level BMR data led to an estimated 1634 (95% CI: 836, 2397) deaths attributable to $NO_2$, while paired with block-level BMR data led to an estimated 1840 (95% CI: 941, 2698) deaths attributable to $NO_2$. However, using county-level $NO_2$ concentrations, instead of 1 km × 1 km $NO_2$ concentrations, results in a > 50% decrease in mortality attributable to $NO_2$ across years. For example, in 2000, using county-level $NO_2$ and county-level BMR data led to 859 (95% CI: 436, 1267) estimated deaths attributable to $NO_2$ (**Figure 4**, **Table S3**).

The mortality attributable to each pollutant per 10,000 residents for the year 2020 is displayed in **Figure 5** at the blockgroup, census tract and county spatial scales. Note, we do not display these values at the block-level because the population data we have is at a coarser resolution than the census block, thus resulting in error when assigning population-levels to each block. We observe spatial differences in the mortality attributable to each pollutant when conducting our analyses at different spatial scales. **Figure S6** displays the top decile of mortality attributable to $PM_{2.5}$ and $NO_2$ per 10,000 residents using data at the block-group, census tract and county-levels for the Denver metropolitan area for the year 2020. There are numerous differences in the hotspots identified at the different spatial scales. For example, when conducting our analysis at the county-level none of the counties in the Denver metropolitan area are in the top-decile of mortality attributable to $PM_{2.5}$, even though several locations in the area show up as hotspots when repeating this analysis at the block-group and censustract scales.

We map the spatial distribution of the % mortality attributable (mortality attributable to each pollutant/all-cause mortality) to each pollutant in **Figure S7** at the block, blockgroup, census tract and county spatial scales for the year 2020. The overall time-series for the % mortality attributable to each pollutant at different spatial scales is also displayed in **Figure 4**. Using block-level data allows for the identification of localized hotspots with a relatively high % of mortality attributable to pollution and high mortality attributable to pollution rates. We also observe that the spatial patterns of % mortality attributable to pollution are similar for analyses at the block- and block-group level. However, these patterns are quite different from those obtained when using census-tract and county-level data, likely due to the averaging out of several local hotspots at these scales. For example, conducting this analysis at the census tract level yields a hotspot of % mortality attributable to pollution (9th decile) in Bent County in South-East Colorado for the year 2020, which is not visible in block-group/block-level analyses (**Figure S7**).

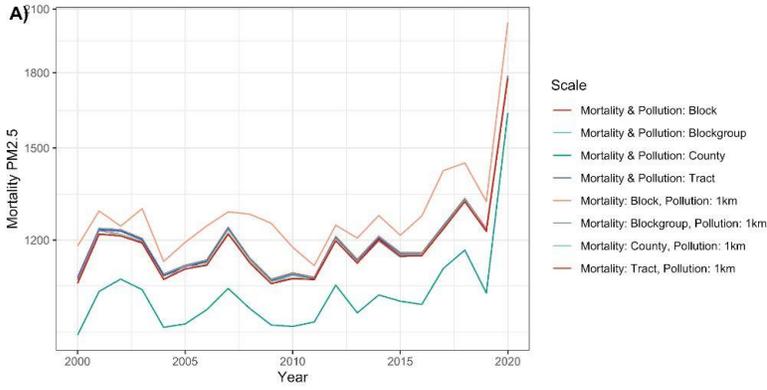
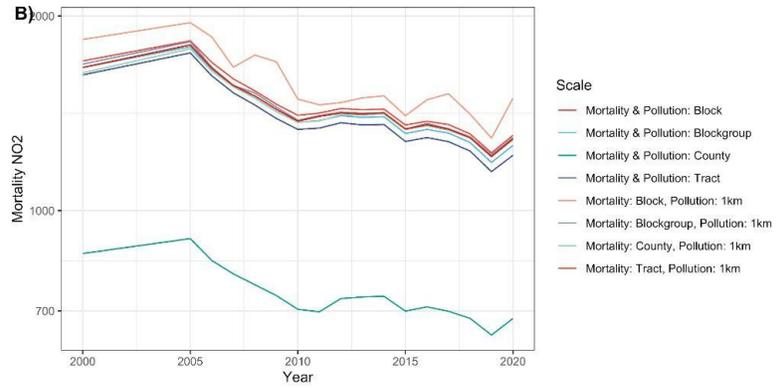
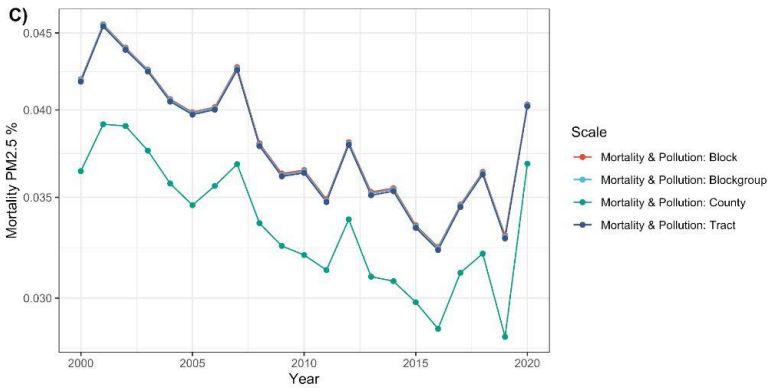
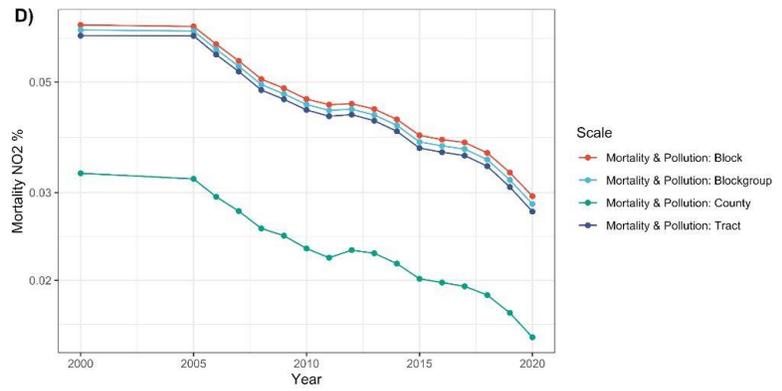
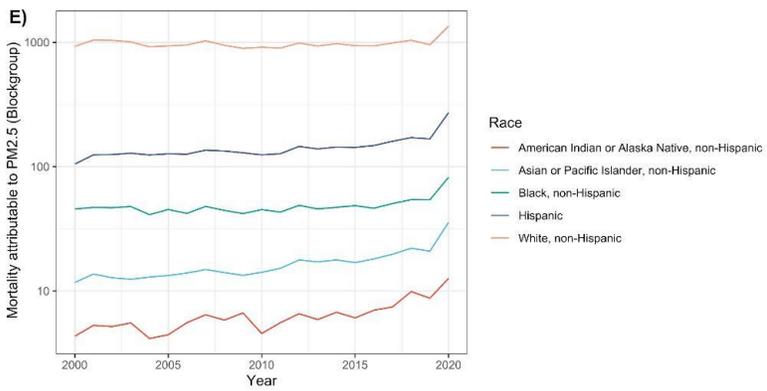
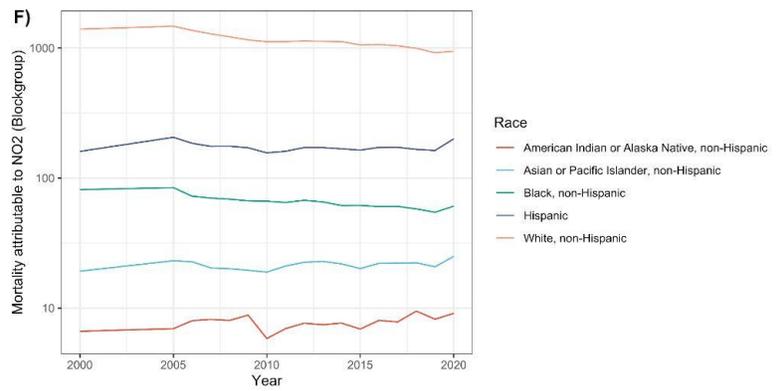
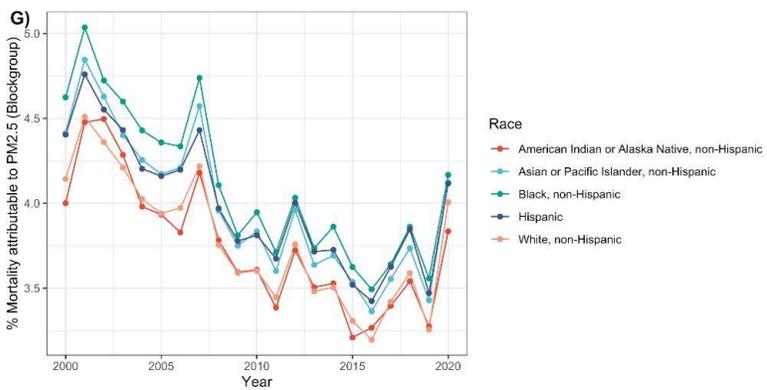
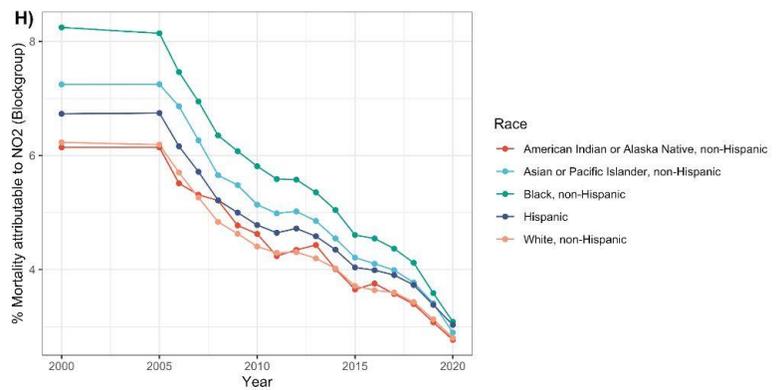

*Figure 4*: Mortality attributable to A) PM$_{2.5}$ and B) NO$_2$, and % mortality attributable (mortality attributable to pollution/all-cause mortality) to C) PM$_{2.5}$ and D) NO$_2$ over time in Colorado using pollution and mortality data at different spatial-scales, Mortality attributable to E) PM$_{2.5}$ and F) NO$_2$ by race using data at the block group-level, % Mortality attributable to G) PM$_{2.5}$ and H) NO$_2$ by race in Colorado over time estimated by summing attributable-mortality estimates produced from using block-group level pollution and BMC data.

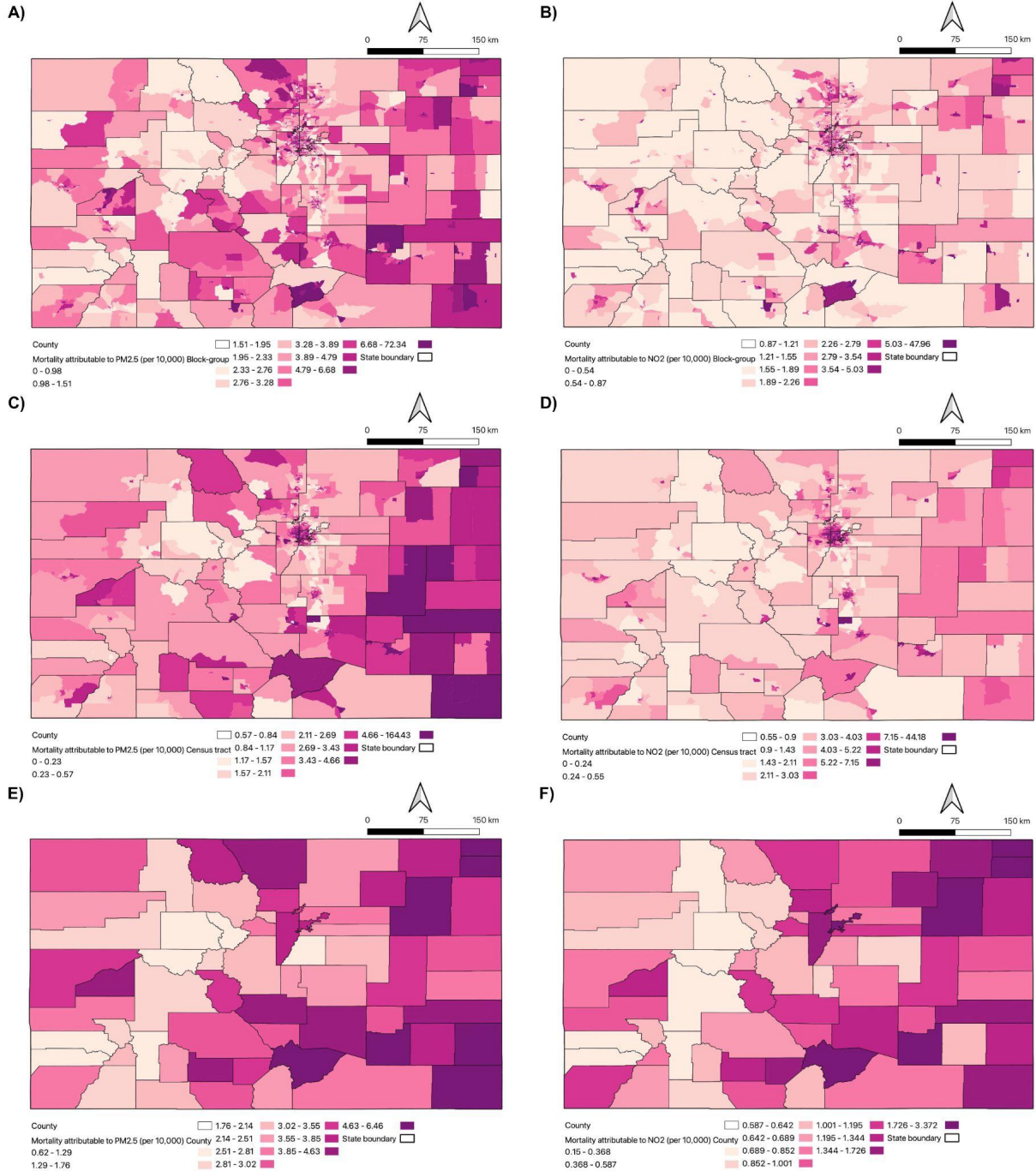

*Figure 5*: Mortality attributable to PM$_{2.5}$ and NO$_2$ per 10,000 residents when using BMC and pollution data at the A) and B) blockgroup, C) and D) census tract, E) and F) county levels, respectively for the year 2020, classified into deciles.

We then repeated this exercise using race-specific mortality data to obtain race-specific mortality attributable to PM$_{2.5}$ and NO$_2$. We observed similar results regarding the impact of spatial scale of pollution and mortality data on the estimated mortality attributable to each pollutant (**Tables S4**, **S5**; **Figure S8**). Plots of the percentage of mortality attributable to each pollutant relative to total mortality by race (**Figure 4**) demonstrates that PM$_{2.5}$ and NO$_2$ burdens non-white Coloradans to a greater degree than white Coloradans. For example, in 2005, for white, non-Hispanic residents, the % of mortality attributable to PM$_{2.5}$ and NO$_2$ were 3.9% and 6.2%, respectively; while for Black residents, these estimates were 4.4 and 8.1%, respectively. Although racial differences persist when evaluating % mortality attributable to NO$_2$ over time, these differences decrease. However, the racial differences in % mortality attributable to PM$_{2.5}$ remain similar over time.

**Figures S9** and **S10** display the mortality attributable to PM$_{2.5}$ and NO$_2$, respectively, per 10,000 residents by racial/ethnic group for the two largest racial/ethnic groups in Colorado: white, non-Hispanic individuals and Hispanic individuals in the year 2020 calculated at the block-group and census tract levels. As with the overall analysis the spatial patterns obtained at the block-group level are different from that at the census tract-level. Moreover, the spatial patterns of mortality attributable to each pollutant are different for the two racial groups due to differences in BMR between the two groups. **Figures S9** and **S10** suggest that using a racial/ethnic-group specific BMR instead of an overall BMR will impact the estimates of mortality attributable to pollution for different racial/ethnic groups.

We then evaluated the impact of using racial/ethnic-specific CRFs on HIA, instead of using a single CRF for the entire population, using pollution and BMC data at the blockgroup-level. Overall, using a single CRF, instead of a racial/ethnic group-specific CRF over-estimates mortality attributable to PM$_{2.5}$ by a factor of ~ 1.3 for the white population (the overall CRF and white-specific CRF for NO$_2$ is the same), and under-estimates mortality attributable to PM$_{2.5}$ and NO$_2$ for the Black population corresponding to factors of ~ 0.4 and 0.5, respectively, every year. Using a Hispanic-specific CRF, instead of the overall CRF results in an underestimation of mortality attributable to PM$_{2.5}$ corresponding to ~ 0.8 and an overestimation of mortality attributable to NO$_2$ corresponding to ~ 2.9, every year (**Figure S11**). Our results thus suggest that accounting for subgroup specific CRFs is important. **Figure S12** displays the % difference in mortality attributable to PM$_{2.5}$ and NO$_2$ $\frac{100 \times (Mortalityy_{Race-specific\ CRF} - Mortalityy_{Single\ CRF})}{Mortalityy_{Single\ CRF}}$ for the year 2020 calculated using a single CRF versus a racial/ethnic group-specific CRF for White and Hispanic individuals. There is little spatial variation in the % difference in mortality attributable to each pollutant based on using a single CRF versus subgroup specific CRFs.

Finally, we evaluated the sensitivity of the HIA to considering mobility patterns when assigning exposures to residents based on home and work-locations (PM$_{2.5,HW}$) to just considering home-locations. Using the latter exposure, instead of the former, results in a decrease in an underestimation of annual mortality attributable to $PM_{2.5}$ by a factor of 0.997 and for $NO_2$ by a factor of 0.980 (**Figure 6**).

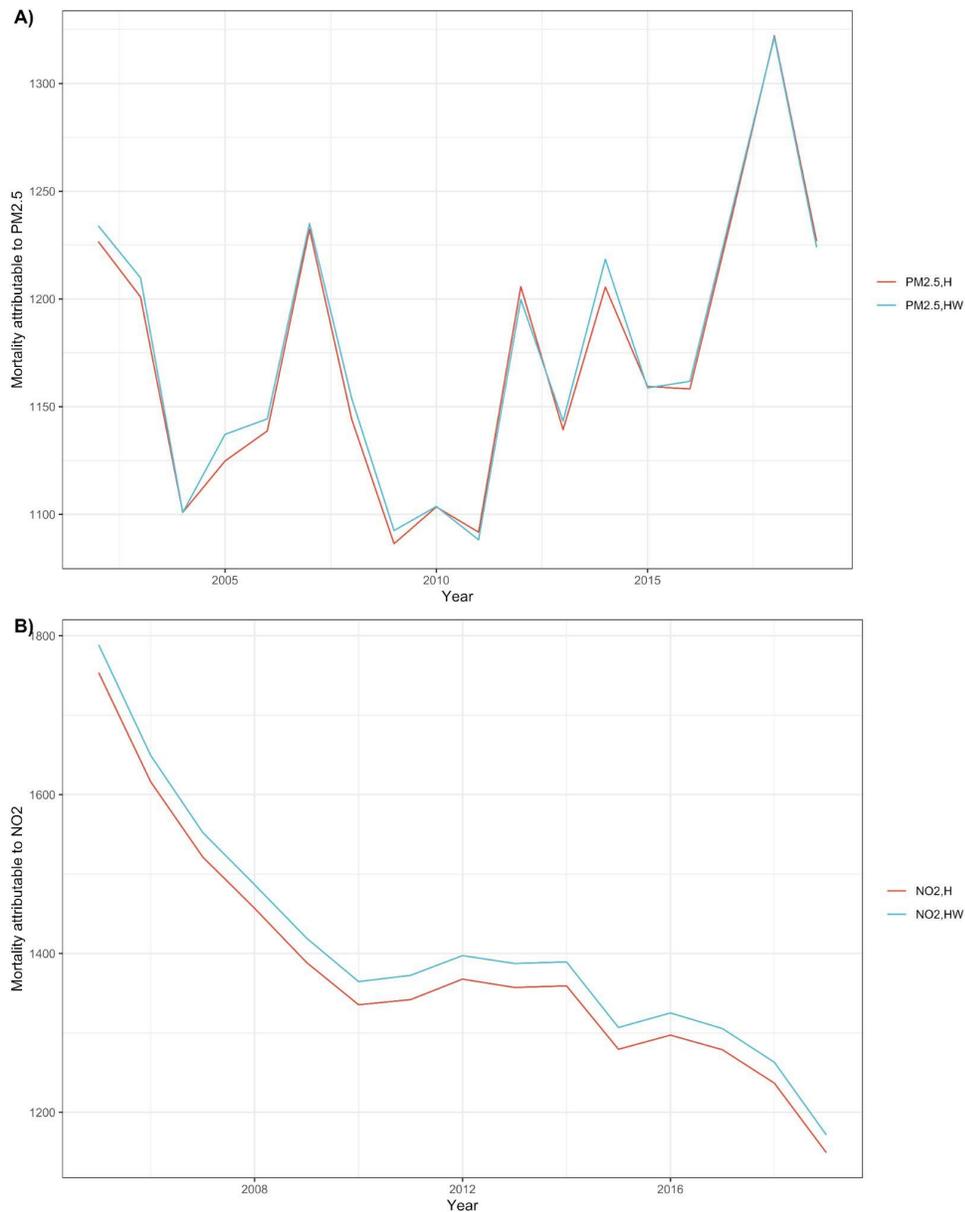

**Figure 6**: *Mortality attributable to A) $PM_{2.5}$ and B) $NO_2$ when assigning exposures based on residential location and residential and work locations.*

# 4 Discussion

At the national scale Boing et al., (2020) found that greatest variation in life-expectancy derived from the United States Small-Area Life Expectancy Estimates Project (US ALEEP) for the time period 2011-2015 was at the census tract geographic scale, when simultaneously considering multiple levels (state, counties, census tracts). In our analysis in Colorado, we observed that proportionally spatial factors accounted for more variance (68%) in BMR than temporal factors (32%). Within the spatial-levels, the largest variance in BMR was observed at the block-level (47.4%). Our work suggests that census-block level data is needed to appropriately evaluate disparities in mortality across Colorado.

Overall, we found that sensitivity of the mortality attributable to pollution to the different modeling choices considered was different for the two pollutants. When varying the spatial scale of the pollution and health data considered, we observed that as $NO_2$ displays more hyperlocal variation than $PM_{2.5}$ (**Figure 1**), the spatial-scale of the pollution considered was a key-driver of the estimates mortality attributable to $NO_2$, but not for $PM_{2.5}$. For example, using county-level estimates, instead of the 1 km × 1 km predictions, resulted in a lower estimate of mortality attributable to $NO_2$ by 50% (and 10% for $PM_{2.5}$) for all of Colorado each year. Other research that has evaluated the impact of the exposure assessment on the health impact of pollution have reported similar findings. For example, in a nation-wide study of the impact of $PM_{2.5}$ on mortality, researchers found that using exposure assessments at a 2 x 2.5° instead of a finer resolution 0.5 x 0.66° resulted in an 8% lower national mortality estimate (Li et al., 2016). Research evaluating the sensitivity of the estimated $NO_2$-attributable pediatric asthma incidence revealed that using an exposure assessment at a 10 $km^2$ and 100 $km^2$ resolution resulted in a lower estimate of pediatric asthma by 6% and 32% respectively across the United States than using the 1 $km^2$ resolved dataset (Mohegh et al., 2020). We note however the generalizability of our results need to be interpreted with caution. For example, the sensitivity of mortality attributable to $PM_{2.5}$ to spatial scale of the exposure could vary depending on the types and spatial patterns of $PM_{2.5}$ sources in specific locales, which might differ from those we observed in Colorado. This is one of the few studies that also tests the sensitivity of the mortality attributable to $PM_{2.5}$ and $NO_2$ to the spatial scale of the health data considered. Using block-level instead of county-level BMR estimates yielded a 10% higher annual mortality attributable to $PM_{2.5}$ and $NO_2$ for all of Colorado.

We observed that using racial/ethnic-specific CRFs instead of an overall CRF resulted in the estimates of mortality attributable to each pollutant every year for all of Colorado for different subpopulations differing by as much as a factor of 2.9 (using an overall-CRF results in a higher mortality attributable to $NO_2$, instead of using a subpopulation specific CRF for Hispanic residents). Note that for both pollutants, using an overall CRF instead of a subpopulation-specific CRF results in a lower estimate of mortality attributable to $PM_{2.5}$ and $NO_2$ for Black residents by 0.4 and 0.5, respectively. Our results, like previous research (Spiller et al., 2021), suggests that especially when evaluating disparities in the health-impacts of pollution, considering racial/ethnic-specific CRFs is important, especially as additional population-specific CRFs are developed and evaluated.

This is the first study that further evaluated the impact of considering home- and work-place based exposures, instead of home-based exposure, alone. We observed that considering mobility patterns resulted in a higher estimate of total mortality attributable to $NO_2$ by a factor of

1.02, while that for $PM_{2.5}$ remained virtually unchanged. This difference is likely because of the greater variation observed in the spatial distribution of $NO_2$ compared to $PM_{2.5}$.

Our analysis has the following limitations:

1) Past research has demonstrated that the spatial scale of the exposure assessment affect HIA calculations and the estimated disparities in health outcomes attributable to pollution. Southerland et al., (2021), for example, found that a higher spatial resolution of the exposure dataset yields a larger burden of pollution. Importantly, we have shown that However, the exposure datasets that we rely on are the most spatially granular publicly available for the entire state.
2) Assigning residents exposure based on home and work-place exposure in our study may not be appropriate as most epidemiologic studies from which we derive CRFs in this study are based on residential exposures, only, and may not be appropriate when evaluating the health impacts of home-work place- based exposures. Therefore, our use of $PM_{2.5,HW}$ exposures should be seen as indicative only, and as a first step to evaluate how mobility patterns can impact HIA calculations. Further, we acknowledge that the LODES datasets are different from the population considered in our study, i.e. the LODES population only consists of working adults. Finally, we note that when calculating $PM_{2.5,HW}$, we assume residents spend ~ 21% of their time at work and apportion annual $PM_{2.5}$, accordingly. However, we do not consider the diurnal variation in $PM_{2.5}$, which can impact this estimation.
3) The main overall CRFs used in these analyses were reported from systematic reviews of studies evaluating associations between all-cause mortality and each of the pollutants considered. The study populations considered in these studies may not be representative of Colorado. However, given the large number of studies considered in the systematic reviews we relied upon, we believe the CRF chosen is applicable to our study setting. We also note that many of the studies considered in the systematic review analysis for $NO_2$, do not control for other pollutants. This could lead to an over-estimation of the impact of $NO_2$ on all-cause mortality. We chose to use the systematic review with the more conservative impact of $NO_2$ in this analysis.
4) The race/ethnicity specific CRFs used in our analyses were derived from single studies relying on the US nation-wide Medicare population that is not representative of the population in Colorado. However, given the dearth of research that reports subpopulation specific CRFs, we relied on these studies.

# Disclaimer

# References

Anenberg, S.C., Mohegh, A., Goldberg, D.L., Kerr, G.H., Brauer, M., Burkart, K., Hystad, P., Larkin, A., Wozniak, S., Lamsal, L., 2022. Long-term trends in urban NO2 concentrations and associated paediatric asthma incidence: estimates from global datasets. Lancet Planet. Health 6, e49–e58. https://doi.org/10.1016/S2542-5196(21)00255-2

Apte, J.S., Messier, K.P., Gani, S., Brauer, M., Kirchstetter, T.W., Lunden, M.M., Marshall, J.D., Portier, C.J., Vermeulen, R.C.H., Hamburg, S.P., 2017. High-Resolution Air Pollution Mapping with Google Street View Cars: Exploiting Big Data. Environ. Sci. Technol. 51, 6999–7008. https://doi.org/10.1021/acs.est.7b00891

Atkinson, Richard.W., Butland, Barbara.K., Anderson, H.Ross., Maynard, Robert.L., 2018. Long-term Concentrations of Nitrogen Dioxide and Mortality. Epidemiol. Camb. Mass 29, 460–472. https://doi.org/10.1097/EDE.0000000000000847

Baston, D., ISciences, L., Baston, M.D., 2022. Package 'exactextractr.' terra 1, 17.

Boing, A.F., Boing, A.C., Cordes, J., Kim, R., Subramanian, S.V., 2020. Quantifying and explaining variation in life expectancy at census tract, county, and state levels in the United States. Proc. Natl. Acad. Sci. 117, 17688–17694. https://doi.org/10.1073/pnas.2003719117

Boing, A.F., deSouza, P., Boing, A.C., Kim, R., Subramanian, S.V., 2022. Air Pollution, Socioeconomic Status, and Age-Specific Mortality Risk in the United States. JAMA Netw. Open 5, e2213540. https://doi.org/10.1001/jamanetworkopen.2022.13540

Castillo, M.D., Kinney, P.L., Southerland, V., Arno, C.A., Crawford, K., van Donkelaar, A., Hammer, M., Martin, R.V., Anenberg, S.C., 2021. Estimating Intra-Urban Inequities in PM2.5-Attributable Health Impacts: A Case Study for Washington, DC. GeoHealth 5, e2021GH000431. https://doi.org/10.1029/2021GH000431

CDC WONDER [WWW Document], n.d. URL https://wonder.cdc.gov/ (accessed 4.13.23).

Chowdhury, S., Dey, S., 2016. Cause-specific premature death from ambient PM2.5 exposure in India: Estimate adjusted for baseline mortality. Environ. Int. 91, 283–290. https://doi.org/10.1016/j.envint.2016.03.004

Colmer, J., Hardman, I., Shimshack, J., Voorheis, J., 2020. Disparities in PM2.5 air pollution in the United States. SCIENCE. https://doi.org/10.1126/science.aaz9353

deSouza, P., Anenberg, S., Makarewicz, C., Shirgaokar, M., Duarte, F., Ratti, C., Durant, J., Kinney, P., Niemeier, D., 2023. Quantifying disparities in air pollution exposures across the United States using home and work addresses. https://doi.org/10.48550/arXiv.2303.12559

DeSouza, P., Anjomshoaa, A., Duarte, F., Kahn, R., Kumar, P., Ratti, C., 2020. Air quality monitoring using mobile low-cost sensors mounted on trash-trucks: Methods development and lessons learned. Sustain. Cities Soc. 60, 102239. https://doi.org/10.1016/j.scs.2020.102239

deSouza, P., Boing, A.F., Kim, R., Subramanian, S., 2022. Associations between ambient PM2.5 – components and age-specific mortality risk in the United States. Environ. Adv. 9, 100289. https://doi.org/10.1016/j.envadv.2022.100289

deSouza, P., Braun, D., Parks, R.M., Schwartz, J., Dominici, F., Kioumourtzoglou, M.-A., 2020. Nationwide Study of Short-term Exposure to Fine Particulate Matter and Cardiovascular Hospitalizations Among Medicaid Enrollees. Epidemiology 32, 6–13. https://doi.org/10.1097/EDE.0000000000001265

deSouza, P.N., Dey, S., Mwenda, K.M., Kim, R., Subramanian, S.V., Kinney, P.L., 2022a. Robust relationship between ambient air pollution and infant mortality in India. Sci. Total Environ. 815, 152755. https://doi.org/10.1016/j.scitotenv.2021.152755

deSouza, P.N., Hammer, M., Anthamatten, P., Kinney, P.L., Kim, R., Subramanian, S.V., Bell, M.L., Mwenda, K.M., 2022b. Impact of air pollution on stunting among children in Africa. Environ. Health 21, 128. https://doi.org/10.1186/s12940-022-00943-y

Di, Q., Wang, Yan, Zanobetti, A., Wang, Yun, Koutrakis, P., Choirat, C., Dominici, F., Schwartz, J.D., 2017. Air Pollution and Mortality in the Medicare Population. N. Engl. J. Med. 376, 2513–2522. https://doi.org/10.1056/NEJMoa1702747

Eum, K.-D., Honda, T.J., Wang, B., Kazemiparkouhi, F., Manjourides, J., Pun, V.C., Pavlu, V., Suh, H., 2022. Long-term nitrogen dioxide exposure and cause-specific mortality in the U.S. Medicare population. Environ. Res. 207, 112154. https://doi.org/10.1016/j.envres.2021.112154

Fann, N., Coffman, E., Timin, B., Kelly, J.T., 2018. The estimated change in the level and distribution of PM2.5-attributable health impacts in the United States: 2005–2014. Environ. Res. 167, 506–514. https://doi.org/10.1016/j.envres.2018.08.018

Fann, N., Fulcher, C.M., Baker, K., 2013. The Recent and Future Health Burden of Air Pollution Apportioned Across US Sectors. Environ. Sci. Technol. 47, 3580–3589. https://doi.org/10.1021/es304831q

Fenech, S., Doherty, R.M., Heaviside, C., Vardoulakis, S., Macintyre, H.L., O'Connor, F.M., 2018. The influence of model spatial resolution on simulated ozone and fine particulate matter for Europe: implications for health impact assessments. Atmospheric Chem. Phys. 18, 5765–5784. https://doi.org/10.5194/acp-18-5765-2018

Hammer, M.S., van Donkelaar, A., Li, C., Lyapustin, A., Sayer, A.M., Hsu, N.C., Levy, R.C., Garay, M.J., Kalashnikova, O.V., Kahn, R.A., Brauer, M., Apte, J.S., Henze, D.K., Zhang, L., Zhang, Q., Ford, B., Pierce, J.R., Martin, R.V., 2020. Global Estimates and Long-Term Trends of Fine Particulate Matter Concentrations (1998–2018). Environ. Sci. Technol. https://doi.org/10.1021/acs.est.0c01764

Huangfu, P., Atkinson, R., 2020. Long-term exposure to NO2 and O3 and all-cause and respiratory mortality: A systematic review and meta-analysis. Environ. Int. 144, 105998. https://doi.org/10.1016/j.envint.2020.105998

Hubbell, B., Fann, N., Levy, J.I., 2009. Methodological considerations in developing local-scale health impact assessments: balancing national, regional, and local data. Air Qual. Atmosphere Health 2, 99–110. https://doi.org/10.1007/s11869-009-0037-z

Institute, H.E., 2022. Systematic Review and Meta-analysis of Selected Health Effects of Long-Term Exposure to Traffic-Related Air Pollution [WWW Document]. Health Eff. Inst. URL https://www.healtheffects.org/publication/systematic-review-and-meta-analysis-selected-health-effects-long-term-exposure-traffic (accessed 7.11.23).

Jiang, X., Yoo, E., 2018. The importance of spatial resolutions of Community Multiscale Air Quality (CMAQ) models on health impact assessment. Sci. Total Environ. 627, 1528–1543. https://doi.org/10.1016/j.scitotenv.2018.01.228

Josey, K.P., Delaney, S.W., Wu, X., Nethery, R.C., DeSouza, P., Braun, D., Dominici, F., 2023. Air Pollution and Mortality at the Intersection of Race and Social Class. N. Engl. J. Med. https://doi.org/10.1056/NEJMsa2300523

Josey, K.P., deSouza, P., Wu, X., Braun, D., Nethery, R., 2022. Estimating a Causal Exposure Response Function with a Continuous Error-Prone Exposure: A Study of Fine Particulate Matter and All-Cause Mortality. J. Agric. Biol. Environ. Stat. https://doi.org/10.1007/s13253-022-00508-z

Khreis, H., Kelly, C., Tate, J., Parslow, R., Lucas, K., Nieuwenhuijsen, M., 2017. Exposure to traffic-related air pollution and risk of development of childhood asthma: A systematic review and meta-analysis. Environ. Int. 100, 1–31. https://doi.org/10.1016/j.envint.2016.11.012

Krewski, D., Jerrett, M., Burnett, R.T., Ma, R., Hughes, E., Shi, Y., Turner, M.C., Pope III, C.A.,

Thurston, G., Calle, E.E., others, 2009. Extended follow-up and spatial analysis of the American Cancer Society study linking particulate air pollution and mortality. Health Effects Institute Boston, MA.

Krupnick, A., Morgenstern, R., 2002. The Future of Benefit-Cost Analyses of the Clean Air Act. Annu. Rev. Public Health 23, 427–448. https://doi.org/10.1146/annurev.publhealth.23.100901.140516

Larkin, A., Geddes, J.A., Martin, R.V., Xiao, Q., Liu, Y., Marshall, J.D., Brauer, M., Hystad, P., 2017. Global Land Use Regression Model for Nitrogen Dioxide Air Pollution. Environ. Sci. Technol. 51, 6957–6964. https://doi.org/10.1021/acs.est.7b01148

Lepeule, J., Laden, F., Dockery, D., Schwartz, J., 2012. Chronic Exposure to Fine Particles and Mortality: An Extended Follow-up of the Harvard Six Cities Study from 1974 to 2009. Environ. Health Perspect. 120, 965–970. https://doi.org/10.1289/ehp.1104660

Li, Y., Henze, D.K., Jack, D., Kinney, P.L., 2016. The influence of air quality model resolution on health impact assessment for fine particulate matter and its components. Air Qual. Atmosphere Health 9, 51–68. https://doi.org/10.1007/s11869-015-0321-z

Mohegh, A., Goldberg, D., Achakulwisut, P., Anenberg, S.C., 2020. Sensitivity of estimated NO2-attributable pediatric asthma incidence to grid resolution and urbanicity. Environ. Res. Lett. 16, 014019. https://doi.org/10.1088/1748-9326/abce25

Parvez, F., Wagstrom, K., 2020. Impact of regional versus local resolution air quality modeling on particulate matter exposure health impact assessment. Air Qual. Atmosphere Health 13, 271–279. https://doi.org/10.1007/s11869-019-00786-6

Pope, C.A., Lefler, J.S., Ezzati, M., Higbee, J.D., Marshall, J.D., Kim, S.-Y., Bechle, M., Gilliat, K.S., Vernon, S.E., Robinson, A.L., Burnett, R.T., n.d. Mortality Risk and Fine Particulate Air Pollution in a Large, Representative Cohort of U.S. Adults. Environ. Health Perspect. 127, 077007. https://doi.org/10.1289/EHP4438

Punger, E.M., West, J.J., 2013. The effect of grid resolution on estimates of the burden of ozone and fine particulate matter on premature mortality in the USA. Air Qual. Atmosphere Health 6, 563–573. https://doi.org/10.1007/s11869-013-0197-8

Siddiqi, S.M., Mingoya-LaFortune, C., Chari, R., Preston, B.L., Gahlon, G., Hernandez, C.C., Huttinger, A., Stephenson, S.R., Madrigano, J., 2022. The Road to Justice40: Organizer and Policymaker Perspectives on the Historical Roots of and Solutions for Environmental Justice Inequities in U.S. Cities. Environ. Justice. https://doi.org/10.1089/env.2022.0038

Southerland, V.A., Anenberg, S.C., Harris, M., Apte, J., Hystad, P., van, D.A., Martin, R.V., Beyers, M., Roy, A., n.d. Assessing the Distribution of Air Pollution Health Risks within Cities: A Neighborhood-Scale Analysis Leveraging High-Resolution Data Sets in the Bay Area, California. Environ. Health Perspect. 129, 037006. https://doi.org/10.1289/EHP7679

Spiller, E., Proville, J., Roy, A., Muller, N.Z., n.d. Mortality Risk from PM2.5: A Comparison of Modeling Approaches to Identify Disparities across Racial/Ethnic Groups in Policy Outcomes. Environ. Health Perspect. 129, 127004. https://doi.org/10.1289/EHP9001

Thurston, G.D., Balmes, J.R., Garcia, E., Gilliland, F.D., Rice, M.B., Schikowski, T., Van Winkle, L.S., Annesi-Maesano, I., Burchard, E.G., Carlsten, C., Harkema, J.R., Khreis, H., Kleeberger, S.R., Kodavanti, U.P., London, S.J., McConnell, R., Peden, D.B., Pinkerton, K.E., Reibman, J., White, C.W., 2020. Outdoor Air Pollution and New-Onset Airway Disease. An Official American Thoracic Society Workshop Report. Ann. Am. Thorac. Soc. 17, 387–398. https://doi.org/10.1513/AnnalsATS.202001-046ST

Tschofen, P., Azevedo, I.L., Muller, N.Z., 2019. Fine particulate matter damages and value added in the US economy. Proc. Natl. Acad. Sci. 116, 19857–19862. https://doi.org/10.1073/pnas.1905030116

Turner, M.C., Jerrett, M., Pope, C.A., Krewski, D., Gapstur, S.M., Diver, W.R., Beckerman, B.S., Marshall, J.D., Su, J., Crouse, D.L., Burnett, R.T., 2016. Long-Term Ozone Exposure and

Mortality in a Large Prospective Study. Am. J. Respir. Crit. Care Med. 193, 1134–1142. https://doi.org/10.1164/rccm.201508-1633OC

Wei, Y., Yazdi, M.D., Di, Q., Requia, W.J., Dominici, F., Zanobetti, A., Schwartz, J., 2021. Emulating causal dose-response relations between air pollutants and mortality in the Medicare population. Environ. Health 20, 53. https://doi.org/10.1186/s12940-021-00742-x

Wu, X., Nethery, R.C., Sabath, M.B., Braun, D., Dominici, F., 2020. Air pollution and COVID-19 mortality in the United States: Strengths and limitations of an ecological regression analysis. Sci. Adv. 6, eabd4049. https://doi.org/10.1126/sciadv.abd4049

# Evaluating the Sensitivity of Pollution-Attributation Mortality to Modeling Choices: A Case Study for Colorado

Priyanka N. deSouza[1,2,3,*], Susan Anenberg[4], Neal Fann[5], Lisa M. McKenzie[6], Elizabeth Chan[5], Ananya Roy[7], Jose L. Jimenez[8,9], William Raich[10], Henry Roman[10], Patrick L. Kinney[11]

1: Department of Urban and Regional Planning, University of Colorado Denver, Denver, CO, USA
2: CU Population Center, University of Colorado Boulder, CO, USA
3: Senseable City Lab, Massachusetts Institute of Technology, USA
4: Miliken Insittute School of Public Health, George Washington University, Washington D.C., USA
5: U.S. Environmental Protection Agency
6: Department of Environmental and Occupational Health, Colorado School of Public Health, University of Colorado Anschutz, Aurora, CO, USA
7: Environmental Defense Fund, NY, USA
8: Cooperative Institute for Research in Environmental Sciences, University of Colorado Boulder, Boulder, CO, USA
9: Department of Chemistry, University of Colorado Boulder, Boulder, CO, USA
10: Industrial Economics, Boston, MA, USA
11: Boston University School of Public Health, MA, USA

*: Corresponding author (priyanka.desouza@ucdenver.edu)

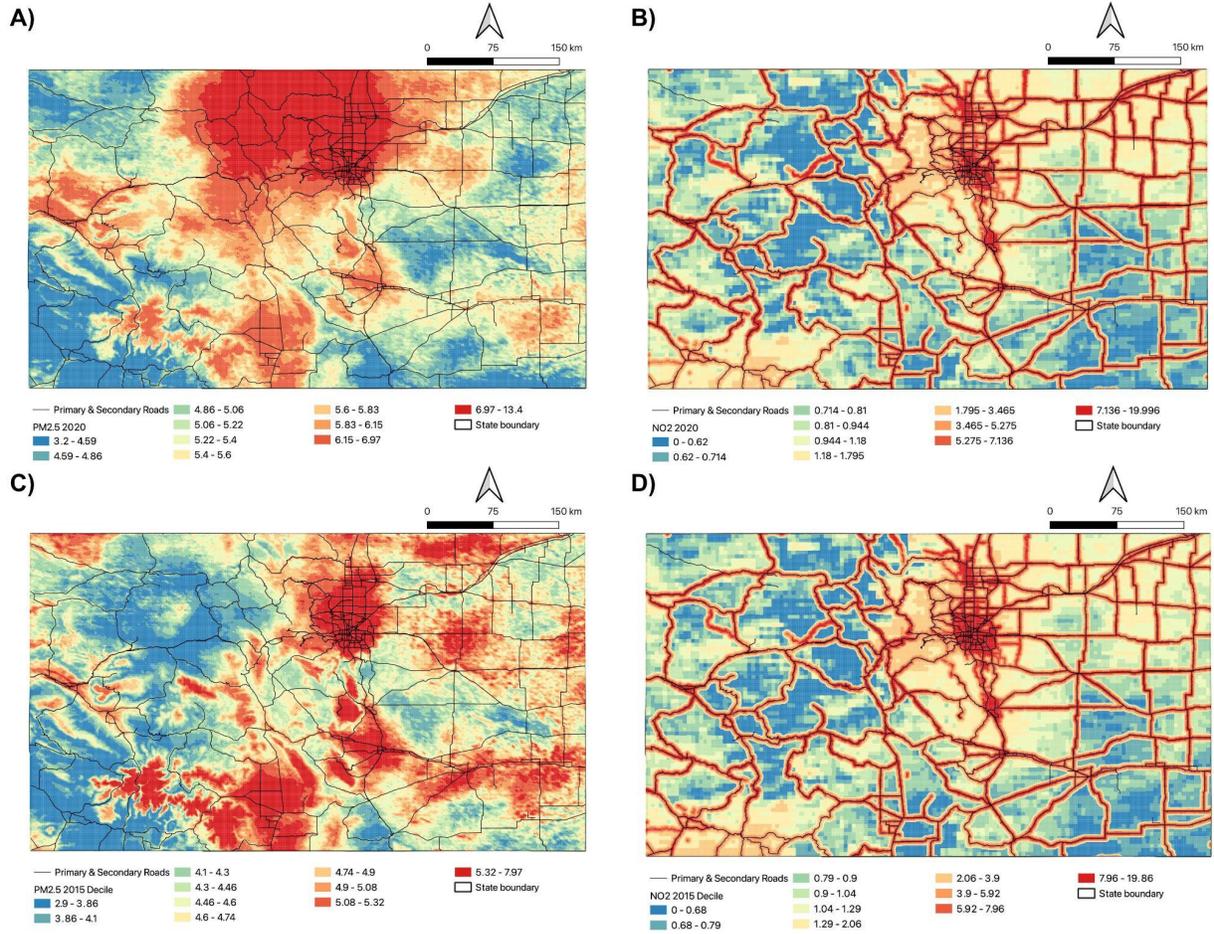

*Figure S1*: Annual-averaged 1 km x 1 km A) PM$_{2.5}$ (µg/m³), and B) NO$_2$ levels (ppbv) over Colorado in 2015 and 2020 classified into deciles.

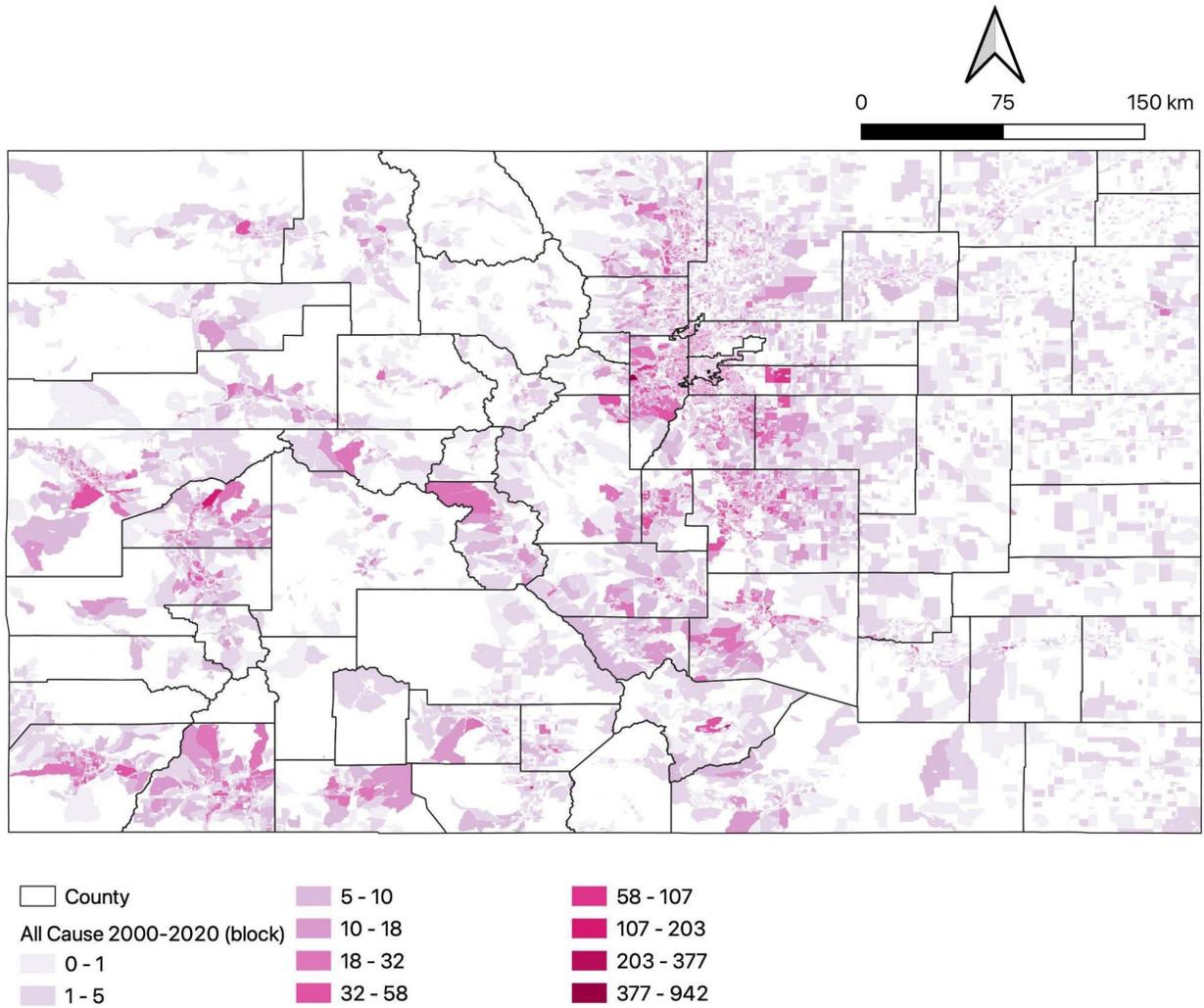

*Figure S2*: Block level, administrative all-cause mortality counts across the years 2000-2020 in Colorado, classified by decile.

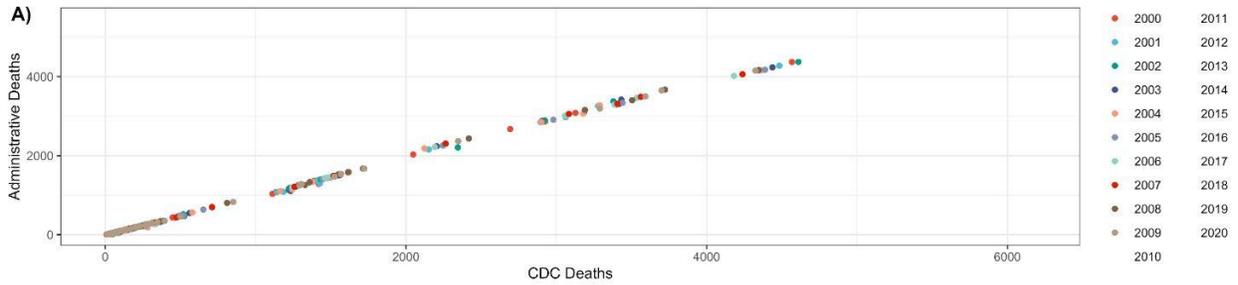
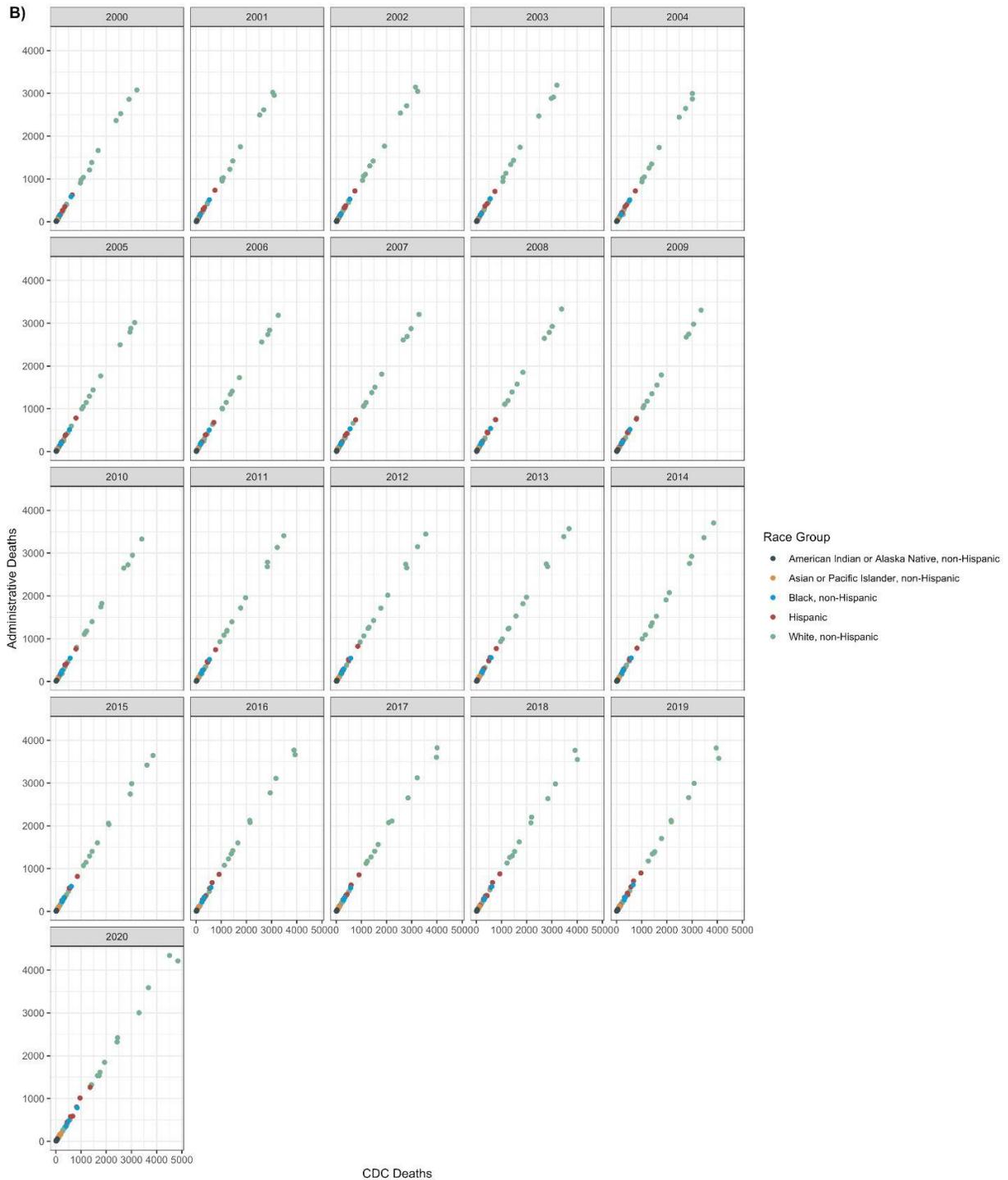

*Figure S3*: County-level baseline mortality from administrative data versus CDC estimates for the years 2000-2020 for A) All-races and B) Different races

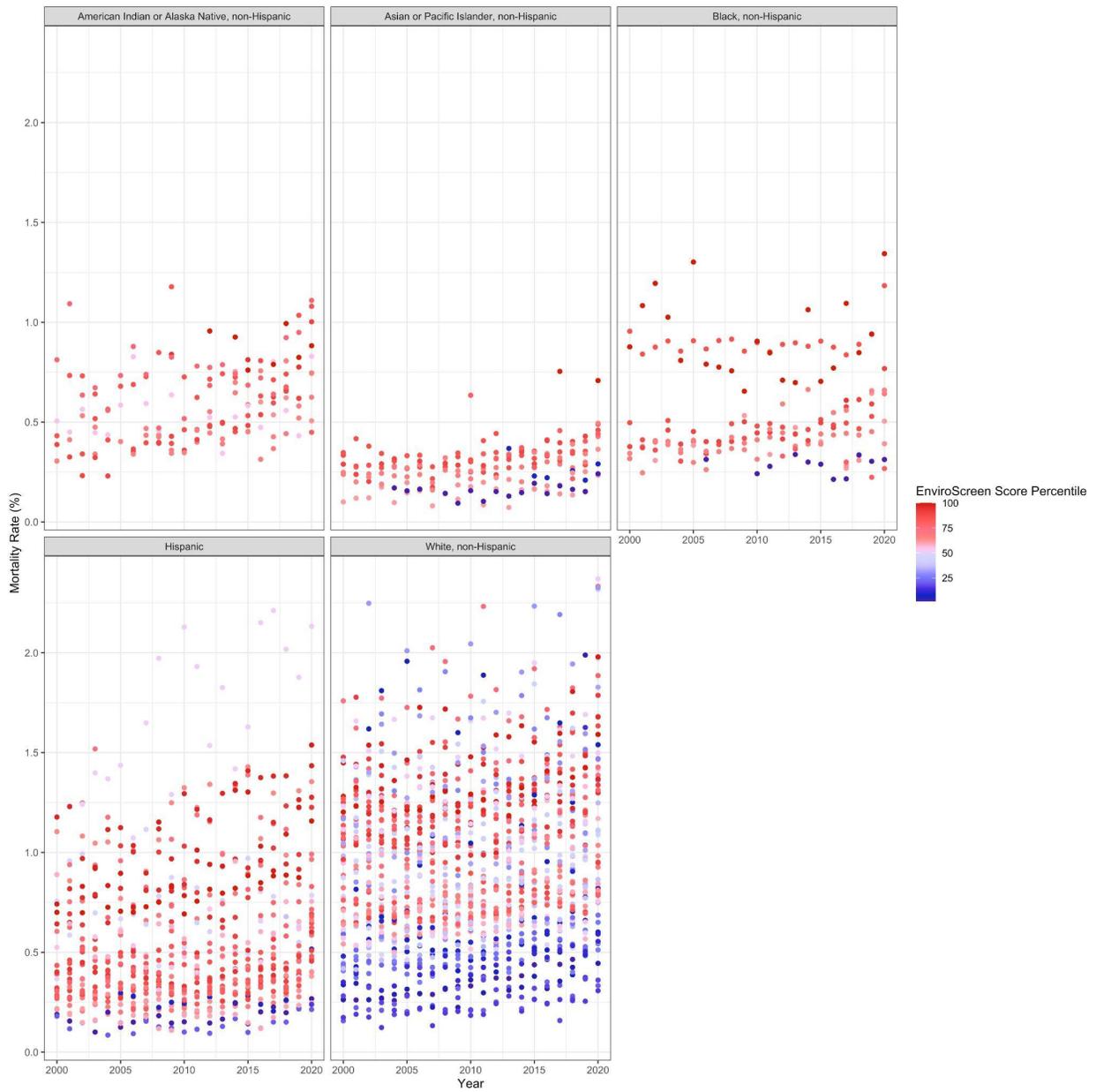

*Figure S4*: County-level, race-specific administrative baseline mortality rate over time, colored by the county-level Colorado Enviroscreen score percentile.

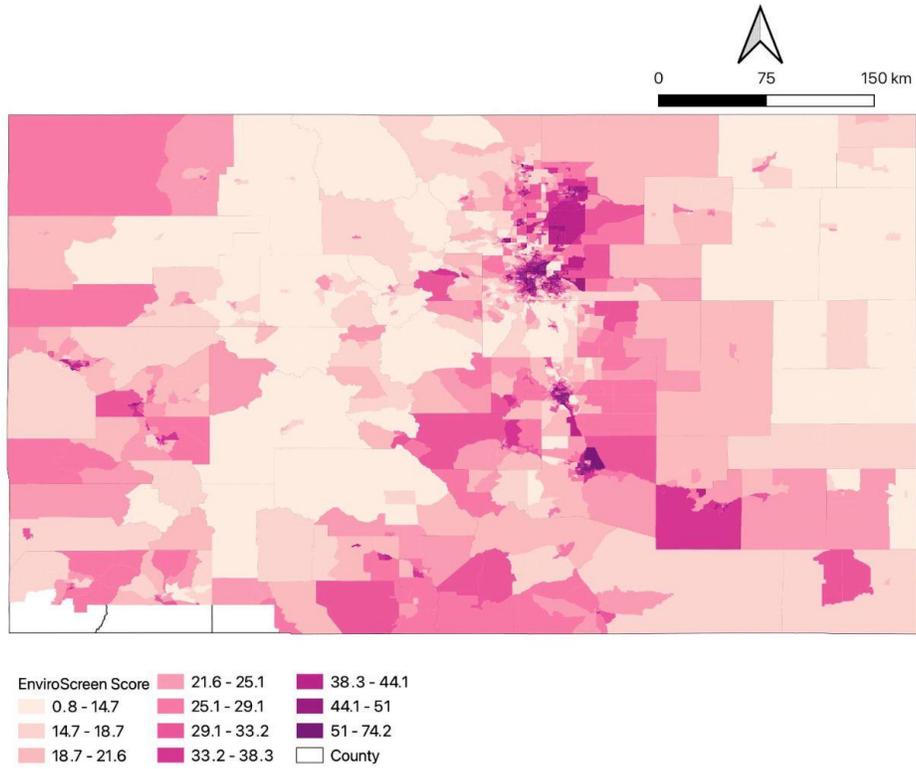
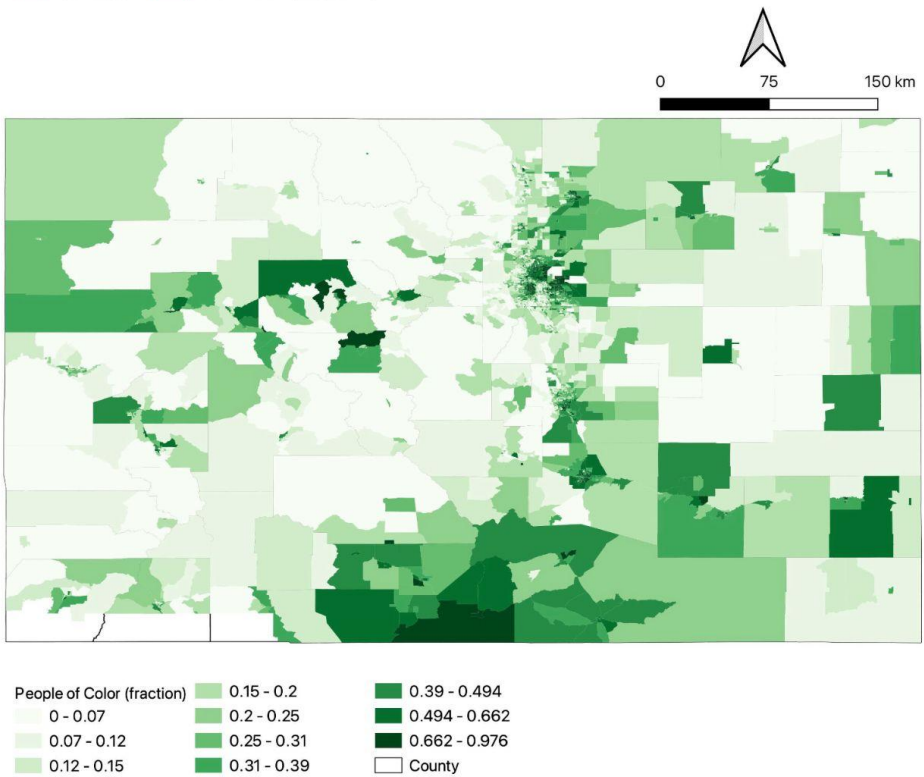

*Figure S5*: A) Blockgroup-level Colorado EnviroScreen score, and B) fraction of non-White Colorado residents for each block group.

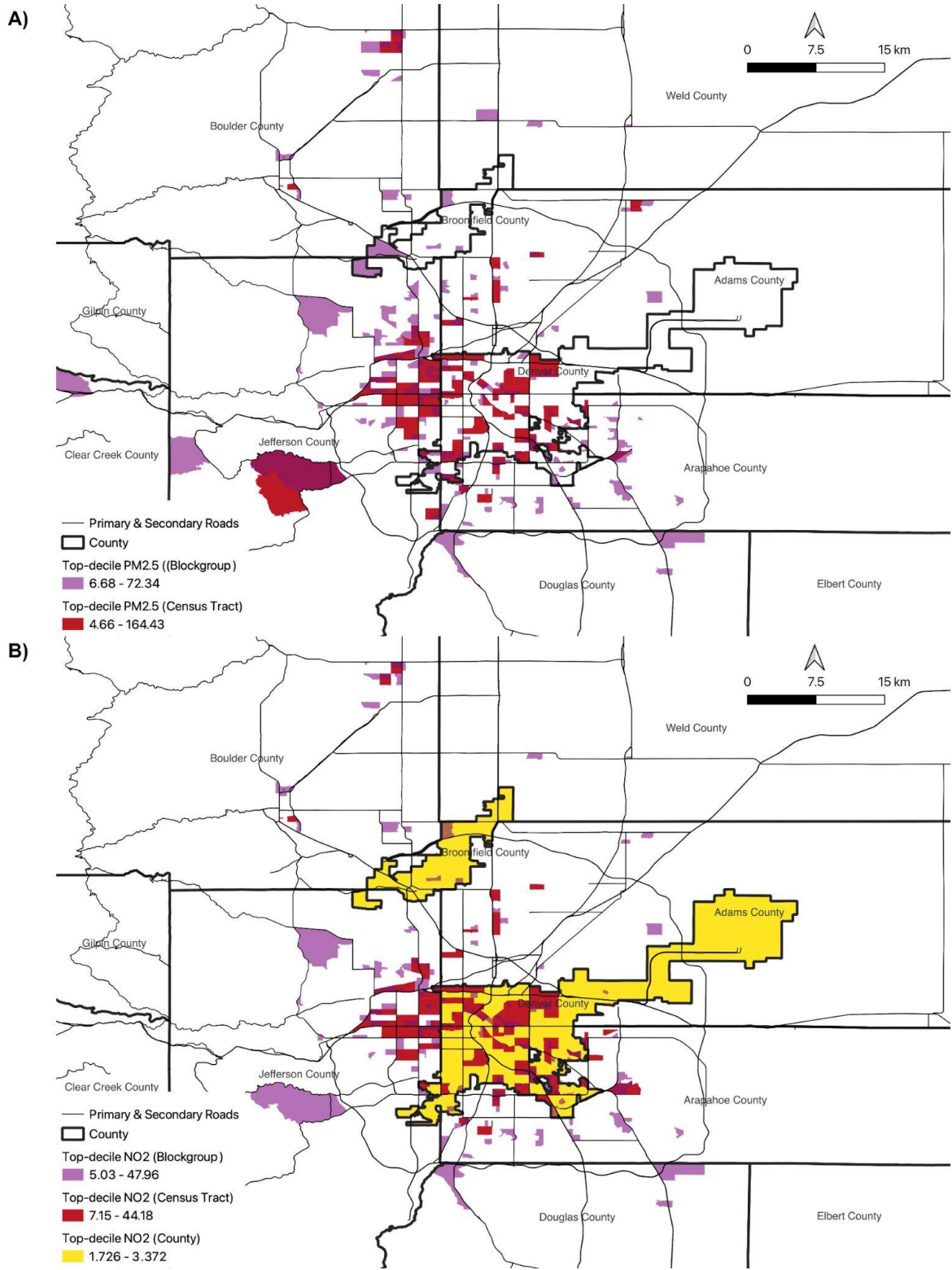

*Figure S6*: Comparison of the top-decile of mortality-attributable to A) PM$_{2.5}$ and B) NO$_2$ per 10,000 residents from using blockgroup and census tract and county- level health and pollution data for the year 2020. Note when conducting the HIA at the county-level for $PM_{2.5}$ data, none of the counties in the Denver metropolitan area are in the top-decile.

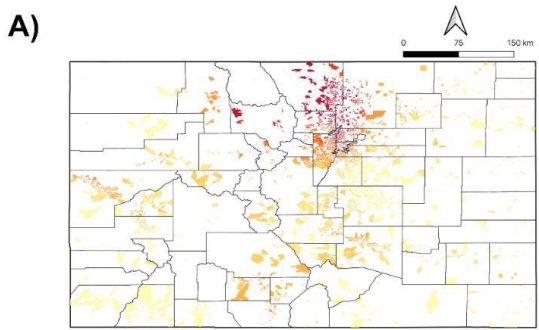
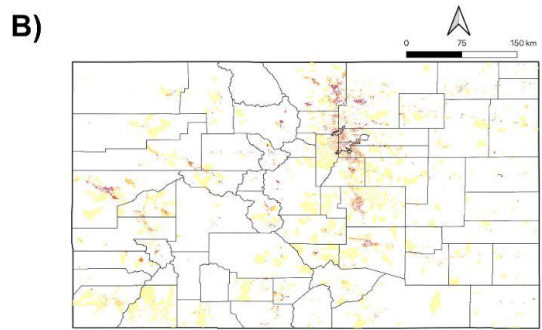
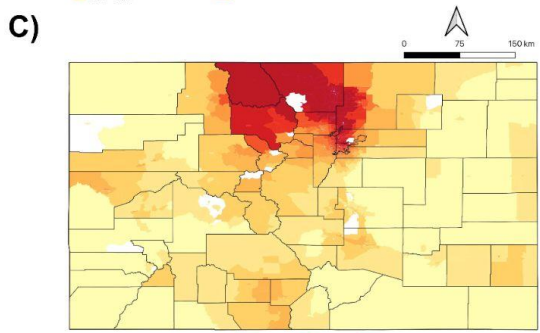
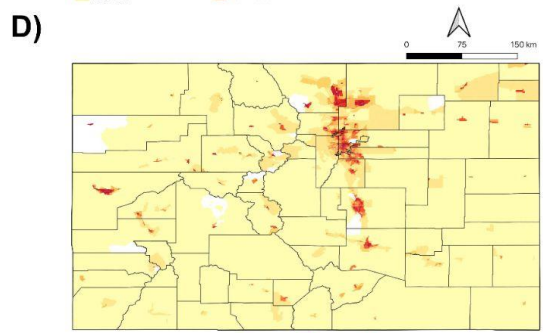
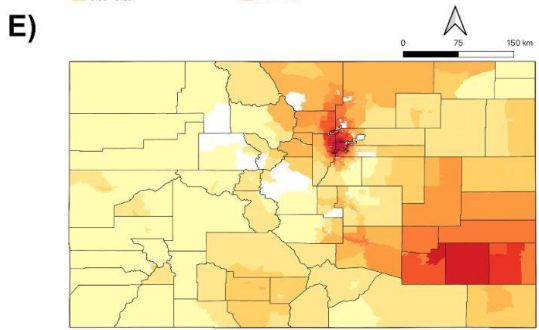
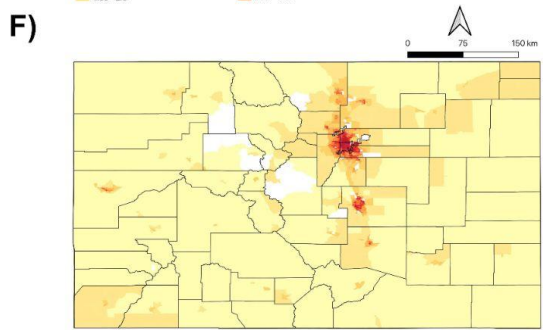
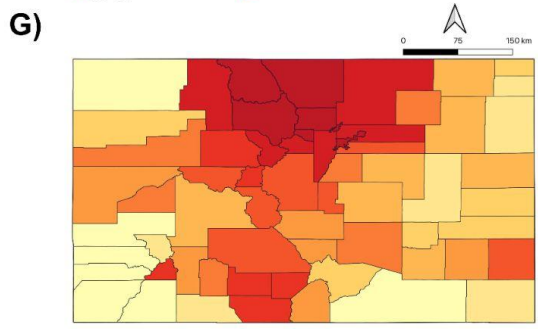
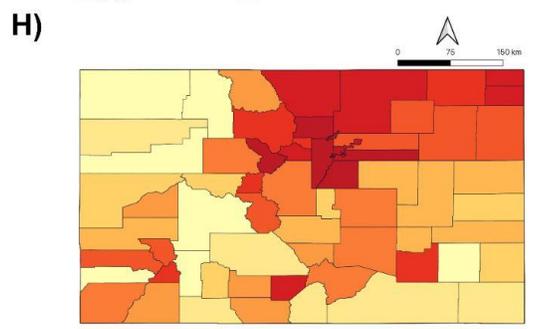

**Figure S7**: % Mortality-attributable to $PM_{2.5}$ and $NO_2$ at the A) and B) block, C) and D) block-group, E) and F) census tract, and G) and H) county levels for the year 2020, classified into deciles.

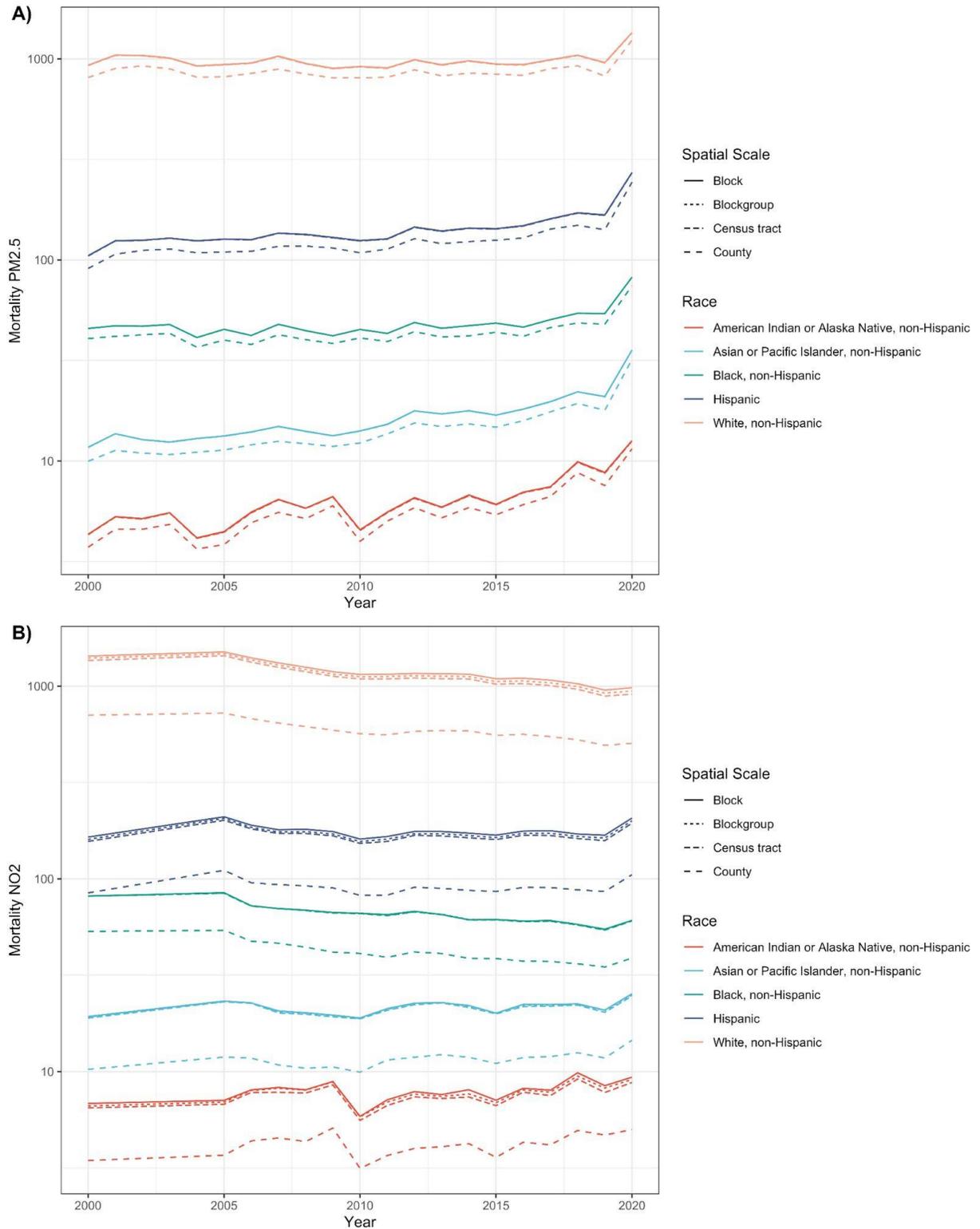

**Figure S8**: Mortality-attributable to A) PM$_{2.5}$ and B) NO$_2$ by race in Colorado over time estimated by summing attributable-mortality estimates produced from using pollution and BMC data at different spatial scales.

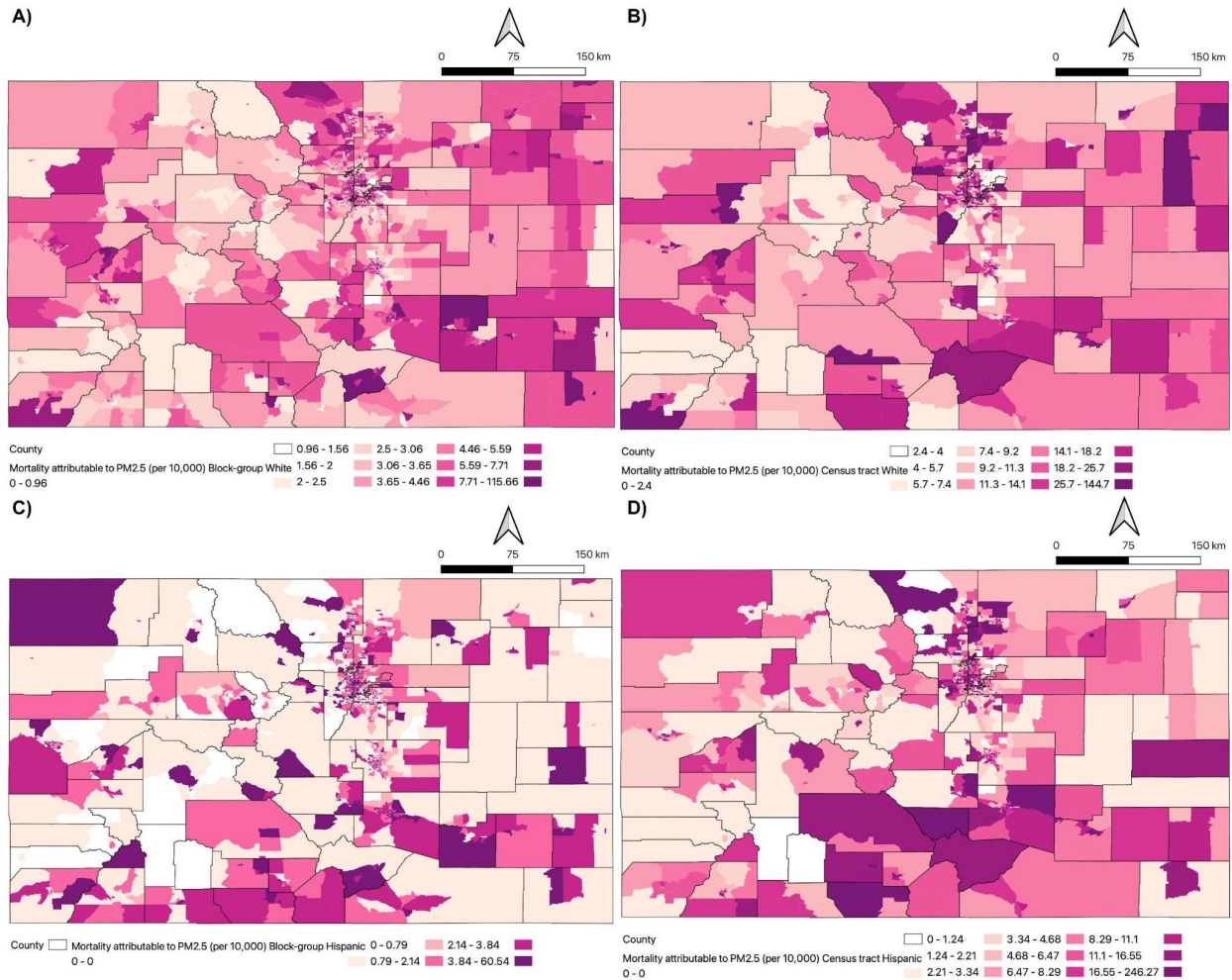

**Figure S9**: Mortality-attributable to PM$_{2.5}$ (per 10,000 residents) using A) and C) blockgroup, and B) and D) census tract level data for A) and B) for White and Hispanic residents for the year 2020, classified into deciles. The number of White and Hispanic residents were obtained from the 2020 decennial census.

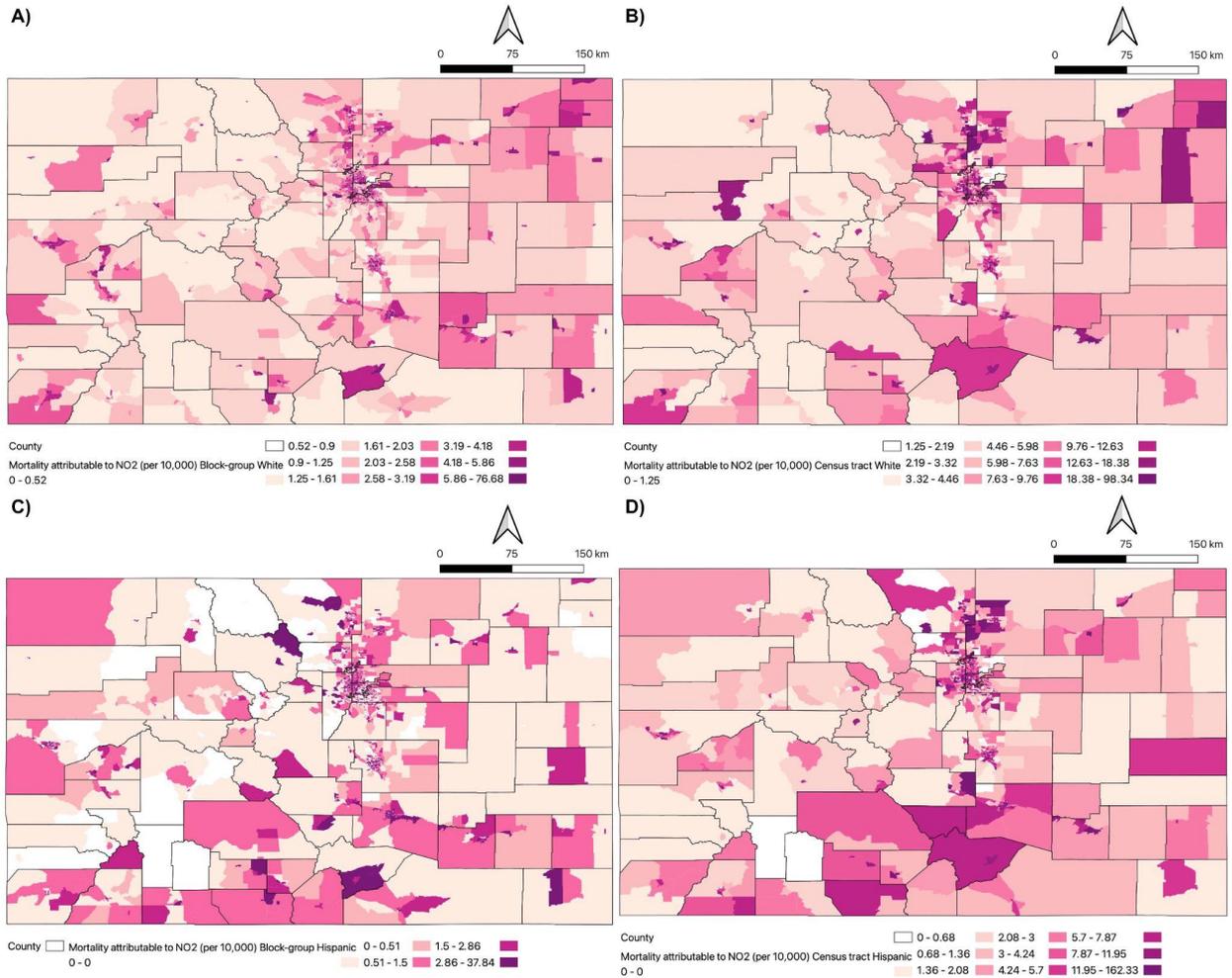

*Figure S10*: Mortality-attributable to NO₂ (per 10,000 residents) using A) and C) blockgroup, and B) and D) census tract level data for A) and B) for White and Hispanic residents for the year 2020, classified into deciles. The number of White and Hispanic residents were obtained from the 2020 decennial census.

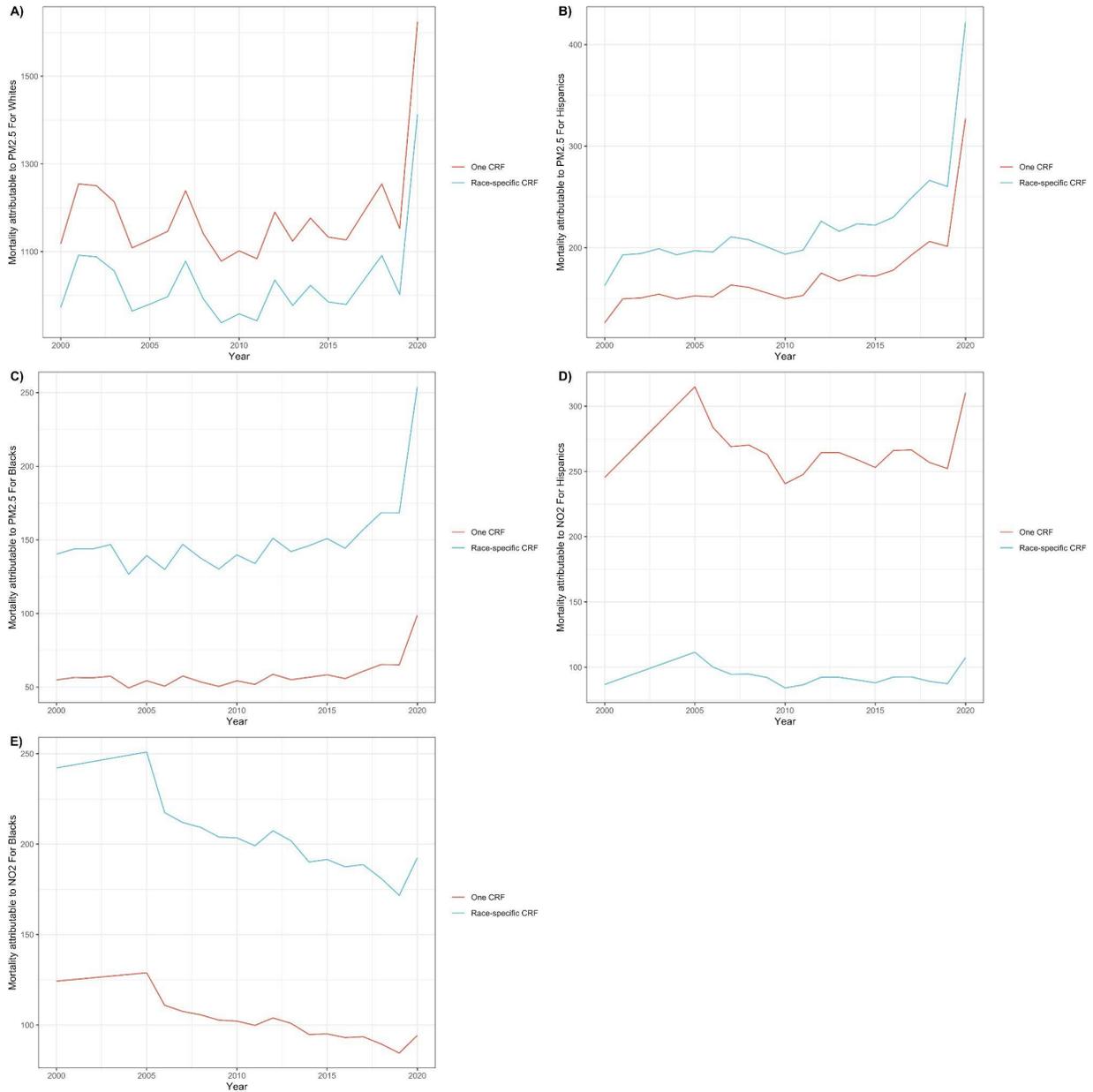

*Figure S11*: Comparing mortality-attributable to $PM_{2.5}$ for A) White, B) Hispanic, C) Black and $NO_2$, for D) Hispanic and E) Black populations over time in Colorado estimated using an overall CRF and racial/ethnic group-specific CRFs (listed in **Table 1**). Note we do not display mortality-attributable to $NO_2$ for the white population because the overall CRF is the same as the CRF for the white, non-Hispanic population. The pollution and health data used in this analysis was at the block-group levels. Overall mortality-attributable to pollution for each year was estimated by summing attributable-mortality estimates at the block-group level over all of Colorado for each year.

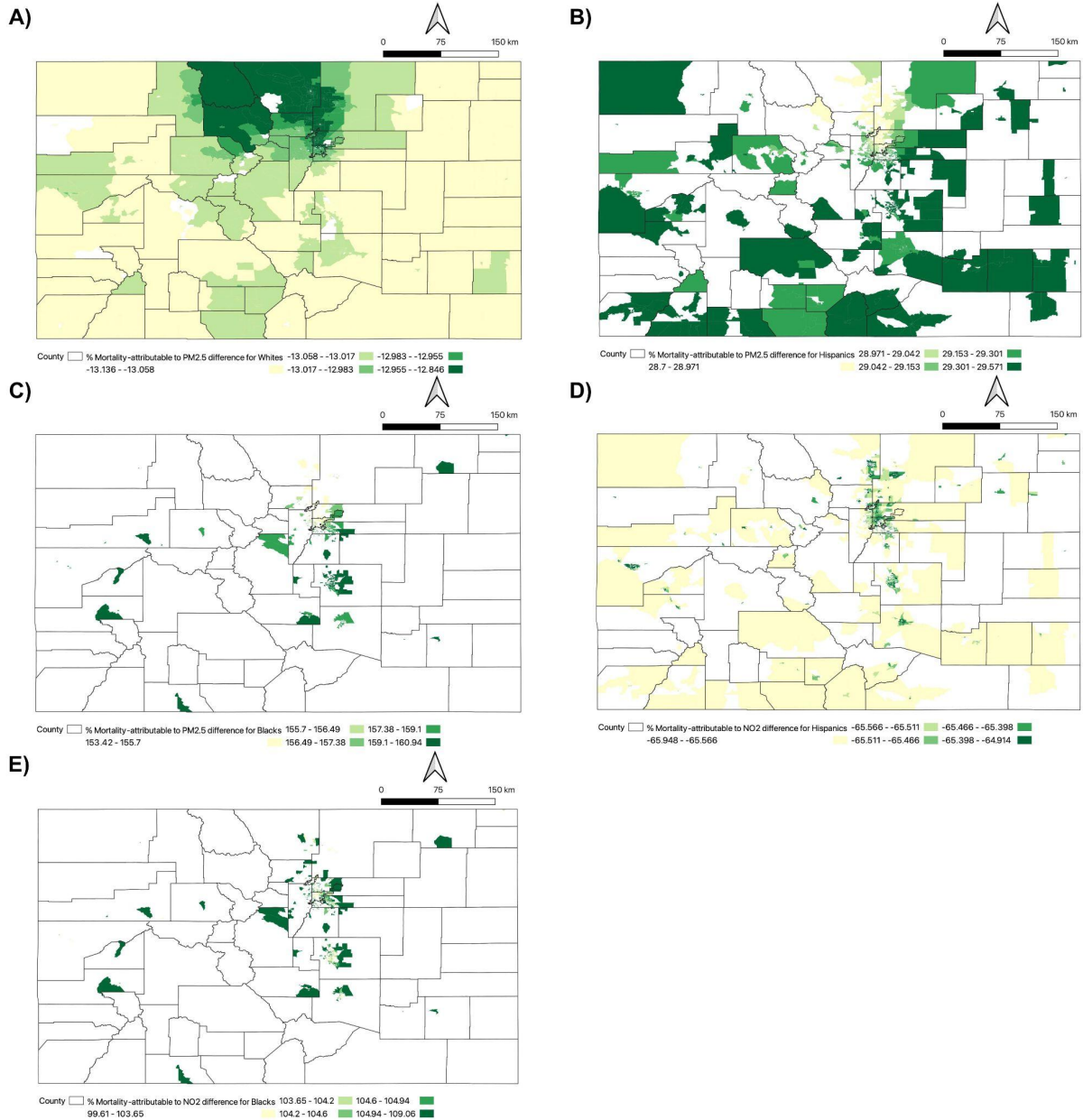

*Figure S12*: % difference in Mortality-attributable to $PM_{2.5}$ and $NO_2$ for 2020 using a single CRF, instead of a racial/ethnic group-specific CRF, calculated using $\frac{100\times (Mortality_{Race-specific\ CRF}-Mortality_{Single\ CRF})}{Mortality_{Single\ CRF}}$ for A) White residents, B) and D) Hispanic residents, C) and E) Black residents, classified into deciles.

# Tables

*Table S1*: BMC and BMR for Colorado derived from the administrative database and the CDC Wonder database at the county-level, as well as mean and population-weighted mean $PM_{2.5}$ and $NO_2$ concentrations for Colorado over time, derived from 1 km × 1 km estimates of the pollutants.

| | | | | Main | | | | H | | HW | |
|---|---|---|---|---|---|---|---|---|---|---|---|
| Year | Administrative data | | CDC Wonder | | $PM_{2.5}$ (µg/m³) | $NO_2$ (ppbv) | $PM_{2.5}$ (µg/m³) | $NO_2$ (ppbv) | $PM_{2.5}$ (µg/m³) | $NO_2$ (ppbv) | $PM_{2.5}$ (µg/m³) | $NO_2$ (ppbv) |
| | BMC | BMR | BMC | BMR | Mean (sd) | | Population-Weighted Mean (sd) | | Population-Weighted Mean (sd) | | Population-Weighted Mean (sd) | |
| 2000 | 26,167 | 0.61% | 27,275 | 0.63% | 5.5 (1.2) | 4.2 (4.2) | 7.1 (1.5) | 17.2 (5.5) | - | - | - | - |
| 2001 | 27,074 | 0.62% | 28,280 | 0.64% | 5.7 (0.9) | - | 7.9 (1.7) | - | - | - | - | - |
| 2002 | 27,969 | 0.63% | 29,199 | 0.65% | 6.1 (0.9) | - | 7.6 (1.2) | - | 7.7 (1.2) | - | 7.8 (1.2) | - |
| 2003 | 28,312 | 0.62% | 29,496 | 0.65% | 5.9 (0.8) | - | 7.3 (1.3) | - | 7.5 (1.3) | - | 7.5 (1.2) | - |
| 2004 | 27,170 | 0.59% | 28,297 | 0.62% | 5.6 (0.8) | - | 7.0 (1.2) | - | 7.1 (1.2) | - | 7.1 (1.1) | - |
| 2005 | 28,318 | 0.61% | 29,616 | 0.64% | 5.4 (0.7) | 3.5 (3.9) | 6.9 (1.3) | 17.6 (6.6) | 7.0 (1.3) | 17.6 (7.1) | 7.1 (1.2) | 18.0 (6.6) |
| 2006 | 28,454 | 0.60% | 29,508 | 0.63% | 5.7 (0.8) | 3.4 (3.7) | 6.9 (1.1) | 16.1 (5.7) | 7.0 (1.1) | 16.0 (6.2) | 7.0 (1.0) | 16.4 (5.8) |
| 2007 | 28,985 | 0.60% | 29,977 | 0.62% | 5.4 (0.8) | 3.3 (3.5) | 7.4 (1.5) | 14.9 (5.1) | 7.5 (1.5) | 14.8 (5.6) | 7.6 (1.4) | 15.2 (5.1) |
| 2008 | 30,223 | 0.62% | 31,262 | 0.64% | 5.3 (0.7) | 3.2 (3.3) | 6.6 (1.1) | 13.7 (4.6) | 6.7 (1.1) | 13.5 (5.1) | 6.7 (1.0) | 13.8 (4.6) |
| 2009 | 30,063 | 0.60% | 31,161 | 0.63% | 5.3 (0.7) | 3.1 (3.2) | 6.2 (0.9) | 13.1 (4.4) | 6.3 (0.9) | 12.8 (4.9) | 6.3 (0.8) | 13.2 (4.5) |
| 2010 | 30,374 | 0.60% | 31,443 | 0.63% | 4.9 (0.7) | 3.0 (3.1) | 6.3 (1.1) | 12.5 (4.0) | 6.3 (1.1) | 12.0 (4.6) | 6.4 (1.0) | 12.4 (4.2) |
| 2011 | 31,419 | 0.61% | 32,550 | 0.64% | 5.1 (0.8) | 2.9 (3.0) | 6.0 (0.9) | 12.1 (3.8) | 6.0 (0.8) | 11.6 (4.4) | 6.0 (0.8) | 11.9 (4.0) |
| 2012 | 31,792 | 0.61% | 33,121 | 0.64% | 5.4 (0.7) | 3.1 (3.1) | 6.6 (1.0) | 12.2 (3.7) | 6.6 (0.9) | 11.8 (4.3) | 6.6 (0.9) | 12.1 (3.9) |
| 2013 | 32,449 | 0.62% | 33,699 | 0.64% | 4.9 (0.8) | 3.1 (3.2) | 6.1 (1.0) | 12.0 (3.5) | 6.1 (1.0) | 11.5 (4.1) | 6.2 (0.9) | 11.8 (3.7) |
| 2014 | 34,121 | 0.64% | 35,218 | 0.66% | 4.7 (0.6) | 3.1 (3.1) | 6.1 (1.1) | 11.4 (3.3) | 6.2 (1.1) | 11.0 (3.8) | 6.3 (1.1) | 11.3 (3.4) |
| 2015 | 34,711 | 0.64% | 36,336 | 0.67% | 4.6 (0.6) | 3.0 (3.0) | 5.8 (1.0) | 10.6 (3.0) | 5.8 (1.0) | 10.3 (3.4) | 5.9 (0.9) | 10.5 (3.0) |
| 2016 | 35,868 | 0.65% | 37,521 | 0.68% | 4.4 (0.6) | 3.0 (3.0) | 5.6 (1.0) | 10.4 (2.9) | 5.7 (1.0) | 10.0 (3.3) | 5.7 (0.9) | 10.3 (3.0) |
| 2017 | 35,922 | 0.65% | 38,055 | 0.68% | 5.0 (0.6) | 2.9 (3.0) | 6.0 (0.9) | 10.2 (2.9) | 6.0 (0.9) | 9.9 (3.2) | 6.1 (0.8) | 10.1 (2.9) |
| 2018 | 36,476 | 0.65% | 38,509 | 0.68% | 5.1 (0.7) | 2.9 (2.9) | 6.3 (1.0) | 9.8 (2.7) | 6.3 (1.0) | 9.4 (3.0) | 6.4 (0.9) | 9.6 (2.7) |
| 2019 | 37,324 | 0.65% | 39,385 | 0.68% | 4.2 (0.7) | 2.9 (2.9) | 5.7 (1.2) | 8.9 (2.3) | 5.8 (1.2) | 8.5 (2.6) | 5.8 (1.1) | 8.7 (2.3) |
| 2020 | 44,315 | 0.77% | 46,898 | 0.81% | 5.7 (1.2) | 2.7 (2.7) | 7.1 (1.3) | 8.0 (2.2) | - | - | - | - |

*Table S2*: BMC and BMR by Race from Administrative and CDC Data

| Year | White non-Hispanic | | | Black, non-Hispanic | | | Hispanic all races | | | Asian, non-Hispanic | | | American Indian non-Hispanic | | |
|---|---|---|---|---|---|---|---|---|---|---|---|---|---|---|---|
| | Administrative BMC | CDC Wonder BMC | CDC Wonder BMR | Administrative BMC | CDC Wonder BMC | CDC Wonder BMR | Administrative BMC | CDC Wonder BMC | CDC Wonder BMR | Administrative BMC | CDC Wonder BMC | CDC Wonder BMR | Administrative BMC | CDC Wonder BMC | CDC Wonder BMR |
| 2000 | 22,412 | 23,330 | 0.72% | 987 | 966 | 0.61% | 2380 | 2447 | 0.37% | 265 | 243 | 0.27% | 108 | 80 | 0.47% |
| 2001 | 23,123 | 24,100 | 0.73% | 934 | 920 | 0.56% | 2615 | 2670 | 0.38% | 282 | 251 | 0.26% | 118 | 102 | 0.58% |
| 2002 | 23,830 | 24,884 | 0.75% | 991 | 986 | 0.58% | 2750 | 2792 | 0.38% | 276 | 245 | 0.25% | 115 | 101 | 0.50% |
| 2003 | 23,952 | 24,951 | 0.75% | 1038 | 1023 | 0.62% | 2894 | 2942 | 0.39% | 282 | 256 | 0.26% | 129 | 111 | 0.48% |
| 2004 | 22,874 | 23,843 | 0.71% | 928 | 917 | 0.54% | 2957 | 2947 | 0.39% | 304 | 283 | 0.26% | 104 | 72 | 0.43% |
| 2005 | 23,760 | 24,887 | 0.74% | 1037 | 1035 | 0.60% | 3050 | 3089 | 0.39% | 319 | 294 | 0.25% | 113 | 77 | 0.60% |
| 2006 | 23,968 | 24,912 | 0.73% | 971 | 962 | 0.54% | 3003 | 2998 | 0.37% | 331 | 316 | 0.27% | 145 | 124 | 0.53% |
| 2007 | 24,411 | 25,329 | 0.73% | 1010 | 1001 | 0.56% | 3065 | 3074 | 0.36% | 325 | 215 | 0.19% | 154 | 109 | 0.57% |
| 2008 | 25,226 | 26,125 | 0.74% | 1083 | 1066 | 0.58% | 3367 | 3369 | 0.39% | 355 | 327 | 0.26% | 154 | 120 | 0.50% |
| 2009 | 24,938 | 25,886 | 0.73% | 1100 | 1093 | 0.58% | 3418 | 3429 | 0.38% | 356 | 340 | 0.25% | 185 | 158 | 0.62% |
| 2010 | 25,397 | 26,358 | 0.74% | 1143 | 1135 | 0.58% | 3266 | 3271 | 0.35% | 368 | 350 | 0.25% | 126 | 90 | 0.49% |
| 2011 | 26,106 | 27,126 | 0.75% | 1161 | 1156 | 0.57% | 3458 | 3466 | 0.37% | 422 | 398 | 0.26% | 164 | 115 | 0.52% |
| 2012 | 26,295 | 27,417 | 0.75% | 1210 | 1196 | 0.59% | 3633 | 3616 | 0.38% | 448 | 441 | 0.28% | 176 | 168 | 0.59% |
| 2013 | 26,797 | 27,878 | 0.75% | 1224 | 1206 | 0.56% | 3739 | 3768 | 0.38% | 471 | 458 | 0.28% | 168 | 138 | 0.53% |
| 2014 | 27,868 | 29,210 | 0.78% | 1219 | 1227 | 0.57% | 3858 | 3890 | 0.38% | 481 | 450 | 0.28% | 191 | 183 | 0.63% |
| 2015 | 28,440 | 29,887 | 0.79% | 1338 | 1345 | 0.60% | 4056 | 4109 | 0.40% | 478 | 507 | 0.28% | 189 | 164 | 0.55% |
| 2016 | 29,256 | 30,790 | 0.80% | 1326 | 1315 | 0.58% | 4315 | 4346 | 0.41% | 538 | 564 | 0.30% | 214 | 178 | 0.60% |
| 2017 | 28,906 | 31,047 | 0.80% | 1386 | 1435 | 0.59% | 4415 | 4527 | 0.42% | 555 | 574 | 0.30% | 219 | 186 | 0.61% |
| 2018 | 29,016 | 31,285 | 0.80% | 1406 | 1477 | 0.60% | 4453 | 4563 | 0.41% | 591 | 604 | 0.30% | 279 | 228 | 0.69% |
| 2019 | 29,393 | 31,673 | 0.80% | 1521 | 1579 | 0.64% | 4815 | 4937 | 0.43% | 609 | 627 | 0.30% | 267 | 215 | 0.70% |
| 2020 | 33,678 | 36,253 | 0.92% | 1971 | 2076 | 0.80% | 6605 | 6733 | 0.57% | 864 | 903 | 0.42% | 329 | 280 | 0.84% |

**Table S3**: Mortality-attributable to $PM_{2.5}$ in Colorado when conducting our analysis at the block, blockgroup, census-tract and county-scales between 2000-2020

| | $PM_{2.5}$ and BMR at the same resolution | | | | $PM_{2.5}$ 1 km $\times$ 1 km resolution, BMR at different spatial resolution | | | |
|---|---|---|---|---|---|---|---|---|
| Year | Block-level | Blockgroup-level | Census tract-level | County-level | Block-level | Blockgroup-level | Census tract-level | County-level |
| 2000 | 1097 (744, 1439) | 1096 (743, 1437) | 1093 (741, 1434) | 953 (646, 1252) | 1183 (802, 1551) | 1088 (738, 1427) | 1080 (732, 1417) | 1080 (732, 1417) |
| 2001 | 1235 (838, 1618) | 1233 (837, 1616) | 1231 (835, 1613) | 1060 (718, 1390) | 1288 (874, 1687) | 1228 (833, 1609) | 1217 (826, 1595) | 1217 (825, 1596) |
| 2002 | 1231 (835, 1614) | 1229 (834, 1612) | 1227 (832, 1608) | 1092 (740, 1433) | 1241 (842, 1627) | 1217 (826, 1596) | 1212 (822, 1589) | 1217 (825, 1596) |
| 2003 | 1205 (817, 1580) | 1203 (816, 1578) | 1201 (814, 1575) | 1064 (721, 1397) | 1295 (878, 1698) | 1196 (810.6, 1568) | 1192 (808, 1563) | 1191 (807, 1562) |
| 2004 | 1105 (749, 1450) | 1103 (748, 1448) | 1101 (746, 1444) | 971 (658, 1275) | 1139 (772, 1495) | 1092 (740, 1432) | 1091 (739, 1431) | 1094 (741, 1435) |
| 2005 | 1129 (765, 1481) | 1127 (764, 1479) | 1125 (762, 1476) | 979 (663, 1286) | 1193 (809, 1565) | 1126 (763, 1478) | 1119 (758, 1468) | 1118 (758, 1467) |
| 2006 | 1143 (775, 1500) | 1141 (773, 1497) | 1139 (772, 1494) | 1014 (686, 1331) | 1243 (844, 1629) | 1130 (766, 1482) | 1129 (765, 1481) | 1134 (768, 1488) |
| 2007 | 1238 (839, 1623) | 1235 (838, 1620) | 1232 (836, 1616) | 1067 (723, 1401) | 1285 (871, 1685) | 1216 (825, 1595) | 1219 (826, 1598) | 1229 (833, 1611) |
| 2008 | 1149 (778, 1508) | 1146 (777, 1505) | 1144 (775, 1502) | 1017 (688, 1336) | 1278 (866, 1677) | 1144 (775, 1501) | 1136 (770, 1491) | 1141 (773, 1498) |
| 2009 | 1091 (739, 1433) | 1089 (737, 1429) | 1086 (736, 1426) | 977 (661, 1283) | 1250 (846, 1641) | 1085 (735, 1424) | 1080 (731, 1417) | 1084 (734, 1423) |
| 2010 | 1108 (751, 1455) | 1106 (749, 1452) | 1103 (747, 1449) | 973 (659, 1279) | 1179 (799, 1548) | 1094 (741, 1436) | 1093 (740, 1435) | 1102 (746, 1446) |
| 2011 | 1096 (742, 1440) | 1094 (741, 1437) | 1092 (739, 1434) | 984 (666, 1293) | 1128 (764, 1481) | 1090 (738, 1432) | 1091 (738, 1432) | 1090 (738, 1431) |
| 2012 | 1210 (820, 1589) | 1208 (818, 1586) | 1206 (817, 1583) | 1076 (728, 1413) | 1244 (843, 1633) | 1199 (812, 1574) | 1198 (812, 1572) | 1203 (815, 1578) |
| 2013 | 1145 (775, 1503) | 1142 (774, 1500) | 1139 (771, 1496) | 1006 (681, 1322) | 1206 (817, 1584) | 1135 (769, 1491) | 1135 (768, 1490) | 1137 (770, 1493) |
| 2014 | 1211 (820, 1590) | 1209 (819, 1587) | 1206 (816, 1583) | 1051 (711, 1381) | 1274 (863, 1672) | 1197 (811, 1572) | 1202 (814, 1578) | 1201 (813, 1577) |
| 2015 | 1164 (788, 1529) | 1162 (787, 1526) | 1159 (785, 1523) | 1035 (700, 1360) | 1214 (822, 1595) | 1152 (780, 1513) | 1155 (782, 1517) | 1157 (783, 1520) |
| 2016 | 1163 (787, 1528) | 1161 (786, 1526) | 1158 (784, 1522) | 1027 (694, 1350) | 1272 (861, 1671) | 1159 (784, 1522) | 1156 (782, 1518) | 1157 (783, 1520) |

| Year | | | | | | | | |
|---|---|---|---|---|---|---|---|---|
| 2017 | 1243 (842, 1633) | 1241 (841, 1630) | 1239 (839, 1627) | 1120 (758, 1472) | 1420 (961, 1865) | 1232 (834, 1617) | 1233 (835, 1619) | 1236 (837, 1623) |
| 2018 | 1327 (899, 1742) | 1325 (897, 1740) | 1322 (895, 1736) | 1171 (793, 1539) | 1446 (980, 1899) | 1320 (894, 1733) | 1317 (892, 1729) | 1320 (894, 1733) |
| 2019 | 1233 (835, 1619) | 1230 (833, 1616) | 1227 (830, 1611) | 1055 (714, 1387) | 1318 (892, 1731) | 1232 (834, 1618) | 1225 (830, 1610) | 1226 (830, 1610) |
| 2020 | 1788 (1212, 2345) | 1785 (1210, 2342) | 1783 (1208, 2339) | 1633 (1106, 2144) | 2032 (1377, 2666) | 1779 (1206, 2334) | 1777 (1205, 2332) | 1781 (1207, 2336) |

*Table S4*: Mortality-attributable to $NO_2$ in Colorado when conducting our analysis at the block, blockgroup, census-tract and county-scales between 2000-2020

| | $NO_2$ | | | | $NO_2$ 1 km × 1 km resolution, BMR at different spatial resolution | | | |
|---|---|---|---|---|---|---|---|---|
| Year | Block-level | Blockgroup-level | Census tract-level | County-level | Block-level | Blockgroup-level | Census tract-level | County-level |
| 2000 | 1704 (872, 3259) | 1665 (852, 3184) | 1622 (830, 3102) | 859 (436, 1663) | 1840 (941, 3519) | 1685 (862, 3222) | 1665 (852, 3184) | 1634 (835, 3127) |
| 2005 | 1831 (937, 3497) | 1793 (918, 3425) | 1754 (898, 3349) | 906 (460, 1752) | 1953 (1000, 3729) | 1825 (934, 3485) | 1805 (924, 3447) | 1780 (911, 3401) |
| 2006 | 1695 (866, 3249) | 1656 (847, 3174) | 1616 (826, 3097) | 837 (425, 1625) | 1857 (949, 3557) | 1666 (851, 3193) | 1660 (848, 3182) | 1644 (840, 3152) |
| 2007 | 1599 (816, 3072) | 1559 (796, 2997) | 1522 (777, 2924) | 799 (405, 1552) | 1667 (850, 3204) | 1562 (797, 3002) | 1562 (797, 3001) | 1555 (793, 2989) |
| 2008 | 1532 (781, 2952) | 1494 (761, 2878) | 1457 (743, 2807) | 768 (390, 1496) | 1740 (887, 3351) | 1521 (775, 2930) | 1504 (767, 2898) | 1493 (761, 2876) |
| 2009 | 1461 (744, 2818) | 1423 (725, 2745) | 1389 (707, 2679) | 739 (375, 1440) | 1699 (865, 3278) | 1446 (737, 2789) | 1435 (731, 2767) | 1423 (725, 2745) |
| 2010 | 1404 (715, 2712) | 1369 (697, 2645) | 1336 (680, 2580) | 704 (357, 1373) | 1487 (757, 5376) | 1380 (702, 2666) | 1376 (700, 2657) | 1370 (697, 2647) |
| 2011 | 1416 (720, 2736) | 1379 (701, 2665) | 1342 (683, 2594) | 698 (353, 1362) | 1458 (742, 2818) | 1403 (714, 2712) | 1399 (712, 2705) | 1376 (700, 2661) |
| 2012 | 1439 (732, 2781) | 1402 (713, 2711) | 1368 (696, 2645) | 731 (370, 1427) | 1469 (747, 2841) | 1421 (723, 2747) | 1416 (720, 2737) | 1407 (716, 2721) |
| 2013 | 1433 (729, 2771) | 1394 (709, 2697) | 1357 (690, 2626) | 736 (372, 1436) | 1494 (760, 2892) | 1415 (720, 2738) | 1411 (718, 2730) | 1406 (715, 2720) |
| 2014 | 1436 (730, 2781) | 1397 (710, 2707) | 1359 (691, 2634) | 737 (373, 1441) | 1505 (765, 2916) | 1415 (719, 2741) | 1417 (720, 2745) | 1411 (717, 2733) |
| 2015 | 1357 (689, 2633) | 1316 (668, 2554) | 1279 (650, 2483) | 699 (354, 1368) | 1402 (712, 2722) | 1336 (679, 2593) | 1338 (679, 2596) | 1335 (678, 2590) |
| 2016 | 1375 (698, 2668) | 1336 (678, 2593) | 1297 (659, 2519) | 710 (359, 1390) | 1484 (753, 2880) | 1365 (693, 2648) | 1358 (689, 2635) | 1353 (687, 2627) |
| 2017 | 1359 (690, 2638) | 1318 (669, 2560) | 1279 (649, 2484) | 699 (353, 1368) | 1517 (770, 2946) | 1339 (680, 2599) | 1337 (679, 2596) | 1332 (676, 2585) |

| 2018 | 1315 (667, 2555) | 1275 (647, 1885) | 1237 (627, 2404) | 682 (345, 1335) | 1411 (716, 2743) | 1301 (660, 2528) | 1297 (658, 2521) | 1297 (658, 2521) |
|---|---|---|---|---|---|---|---|---|
| 2019 | 1229 (623, 2392) | 1188 (602, 2312) | 1149 (582, 2238) | 642 (324, 1259) | 1295 (656, 2521) | 1220 (618, 2376) | 1213 (615, 2361) | 1211 (614, 2358) |
| 2020 | 1308 (662, 2595) | 1262 (639, 2461) | 1218 (617, 2377) | 681 (344, 1338) | 1492 (755, 2909) | 1297 (657, 2529) | 1291 (654, 2518) | 1287 (652, 2510) |

**Table S5**: Race-specific mortality attributable to $PM_{2.5}$ when conducting our analysis at the block, blockgroup, census-tract and county-scales between 2000-2020

| Year | White | | | | Black | | | | Hispanic | | | |
|---|---|---|---|---|---|---|---|---|---|---|---|---|
| | Block-level | Blockgroup-level | Census tract-level | County-level | Block-level | Blockgroup-level | Census tract-level | County-level | Block-level | Blockgroup-level | Census tract-level | County-level |
| 2000 | 930 (630, 1220) | 929 (630, 1218) | 926 (628, 1215) | 808 (547, 1061) | 46 (31, 60) | 46 (31, 60) | 46 (31, 60) | 41 (28, 53) | 105 (71, 138) | 105 (71, 137) | 105 (71, 137) | 91 (61, 119) |
| 2001 | 1044 (708, 1368) | 1043 (707, 1367) | 1040 (706, 1364) | 895 (607, 1175) | 47 (32, 62) | 47 (32, 62) | 47 (32, 62) | 42 (28, 55) | 125 (85, 163) | 124 (85, 163) | 124 (84, 163) | 106.8 (72.4, 140.1) |
| 2002 | 1040 (706, 1364) | 1039 (705, 1362) | 1037 (703, 1359) | 922 (625, 1210) | 47 (32, 61) | 47 (32, 61) | 46.8 (31.7, 61.3) | 42.4 (28.7, 55.6) | 125 (85, 164) | 125 (85, 164) | 125 (85, 164) | 111 (76, 146) |
| 2003 | 1010 (685, 1324) | 1008 (684, 1323) | 1006 (682, 1320) | 892 (604, 1170) | 48 (32, 63) | 48 (32, 63) | 48 (32, 63) | 43 (29, 57) | 128 (87, 168) | 128 (87, 168) | 128 (87, 168) | 113 (77, 149) |
| 2004 | 922 (625, 1210) | 921 (624, 1208) | 919 (623, 1205) | 811 (549, 1065) | 41 (28, 54) | 41 (28, 54) | 41 (28, 54) | 37 (25, 48) | 124 (84, 163) | 124 (84, 163) | 124 (84, 163) | 109 (74, 143) |
| 2005 | 938 (635, 1230) | 936 (634, 1228) | 934 (633, 1225) | 814 (551, 1068) | 45 (31, 59) | 45 (32, 59) | 45 (31, 59) | 40 (27, 52) | 127 (86, 167) | 127 (86, 166) | 127 (86, 166) | 110 (74, 144) |
| 2006 | 954 (647, 1252) | 952 (645, 1249) | 950 (644, 1246) | 847 (573, 1112) | 42 (29, 55) | 42 (29, 55) | 42 (29, 55) | 38 (26, 50) | 126 (86, 166) | 126 (85, 165) | 126 (85, 165) | 110 (75, 145) |
| 2007 | 1032 (700, 1353) | 1030 (698, 1350) | 1027 (696, 1347) | 889 (602, 1167) | 48 (33, 63) | 48 (33, 63) | 48 (33, 63) | 43 (29, 56) | 136 (92, 178) | 136 (92, 178) | 136 (92, 178) | 117 (79, 153) |
| 2008 | 949 (643, 1246) | 947 (642, 1243) | 945 (640, 1240) | 841 (569, 1105) | 45 (30, 58) | 45 (30, 58) | 45 (30, 58) | 40 (27, 53) | 134 (91, 176) | 134 (91, 175) | 133 (90, 175) | 117 (79, 154) |
| 2009 | 897 (608, 1178) | 895 (606, 1175) | 893 (605, 1173) | 804 (544, 1056) | 42 (28, 55) | 42 (28, 55) | 42 (28, 55) | 38 (26, 50) | 129 (88, 170) | 129 (88, 170) | 129 (87, 169) | 115 (78, 151) |
| 2010 | 917 (621, 1204) | 915 (620, 1201) | 913 (618, 1198) | 805 (545, 1058) | 45 (31, 59) | 45 (31, 59) | 45 (31, 59) | 42 (28, 54) | 125 (85, 164) | 125 (84, 163) | 124 (84, 163) | 109 (73, 143) |
| 2011 | 902 (610, 1184) | 900 (609, 1181) | 897 (608, 1179) | 809 (548, 1064) | 43 (29, 57) | 43 (29, 57) | 43 (29, 57) | 39 (27, 53) | 127 (86, 167) | 127 (86, 167) | 127 (86, 166) | 113 (77, 149) |
| 2012 | 990 (671, | 988 (670, | 986 (668, | 882 (597, | 49 (33, 64) | 49 (33, 64) | 49 (33, 64) | 44 (30, 58) | 146 (99, 191) | 145 (99, 191) | 145 (98, 190) | 128 (86, 167) |

| Year | | | | | | | | | | | | |
|---|---|---|---|---|---|---|---|---|---|---|---|---|
| | 1300) | 1297) | 1295) | 1159) | | | | | | | | |
| 2013 | 935 (633, 1228) | 933 (632, 1225) | 930 (630, 1222) | 823 (557, 1081) | 46 (31, 60) | 46 (31, 60) | 46 (31, 60) | 41 (28, 54) | 139 (94, 183) | 139 (94, 182) | 139 (94, 182) | 120 (82, 158) |
| 2014 | 979 (663, 1285) | 977 (661, 1282) | 974 (660, 1279) | 849 (575, 1116) | 47 (32, 62) | 47 (32, 62) | 47 (32, 62) | 43 (28, 55) | 144 (98, 189) | 144 (97, 189) | 143 (97, 188) | 123 (84, 162) |
| 2015 | 942 (638, 1238) | 941 (637, 1235) | 938 (635, 1232) | 839 (568, 1103) | 49 (33, 64) | 49 (33, 64) | 49 (33, 64) | 44 (30, 57) | 143 (97, 188) | 143 (97, 188) | 143 (97, 187) | 125 (85, 164) |
| 2016 | 937 (634, 1231) | 935 (633, 1228) | 932 (631, 1225) | 828 (560, 1089) | 46 (31, 61) | 46 (31, 61) | 46 (31, 61) | 42 (28, 55) | 148 (100, 194) | 148 (100, 194) | 147 (100, 194) | 129 (87, 169) |
| 2017 | 990 (670, 1301) | 989 (669, 1299) | 986 (668, 1296) | 894 (605, 1175) | 51 (34, 66) | 51 (34, 66) | 51 (34, 66) | 46 (31, 61) | 160 (109, 210) | 160 (108, 210) | 160 (108, 210) | 142 (96, 187) |
| 2018 | 1044 (707, 1370) | 1042 (705, 1368) | 1039 (704, 1365) | 923 (624, 1212) | 54 (37, 71) | 54 (37, 71) | 54 (37, 71) | 49 (33, 64) | 172 (116, 225) | 171 (116, 225) | 171 (116, 224) | 149 (101, 195) |
| 2019 | 959 (649, 1260) | 957 (648, 1257) | 954 (646, 1253) | 821 (555, 1079) | 54 (30, 86) | 54 (37, 71) | 54 (37, 71) | 48 (32, 63) | 168 (113, 220) | 167 (113, 220) | 167 (113, 219) | 141 (96, 186) |
| 2020 | 1351 (916, 1773) | 1349 (914, 1770) | 1347 (913, 1767) | 1240 (840, 1628) | 82 (56, 108) | 82 (56, 108) | 82 (56, 108) | 74 (50, 98) | 272 (185, 357) | 272 (184, 357) | 271 (184, 356) | 244 (165, 320) |

**Table S6**: Race-specific mortality attributable to $NO_2$ when conducting our analysis at the block, blockgroup, census-tract and county-scales between 2000-2020

| Year | White | | | | Black | | | | Hispanic | | | |
|---|---|---|---|---|---|---|---|---|---|---|---|---|
| | Block-level | Blockgroup-level | Census tract-level | County-level | Block-level | Blockgroup-level | Census tract-level | County-level | Block-level | Blockgroup-level | Census tract-level | County-level |
| 2000 | 1431 (732, 2738) | 1396 (714, 2672) | 1357 (694, 2598) | 706 (359, 1370) | 81 (42, 154) | 81 (42, 154) | 81 (42, 154) | 53 (27, 102) | 165 (84, 314) | 160 (82, 306) | 156 (80, 298) | 85 (43, 163) |
| 2005 | 1504 (770, 2875) | 1471 (753, 2811) | 1437 (735, 2745) | 724 (368, 1403) | 85 (44, 161) | 84 (43, 160) | 84 (43, 160) | 54 (28, 104) | 210 (108, 400) | 206 (105, 392) | 201 (103, 383) | 111 (56, 213) |
| 2006 | 1400 (715, 2685) | 1366 (698, 2620) | 1330 (680, 2551) | 677 (344, 1315) | 73 (37, 138) | 72 (37, 138) | 72 (37, 138) | 48 (24, 91) | 190 (97, 363) | 185 (95, 354) | 182 (93, 347) | 95 (49, 185) |
| 2007 | 1319 (673, 2536) | 1285 (655, 2470) | 1251 (638, 2405) | 643 (326, 1251) | 70 (36, 134) | 70 (36, 134) | 70 (36, 134) | 46 (24, 89) | 180 (92, 345) | 175 (89, 336) | 172 (88, 330) | 93 (47, 181) |
| 2008 | 1253 (638, 2414) | 1220 (622, 2351) | 1187 (605, 2289) | 617 (312, 1201) | 69 (35, 132) | 69 (35, 132) | 68 (35, 131) | 44 (22, 85) | 181 (92, 348) | 176 (90, 338) | 172 (88, 331) | 92 (47, 179) |
| 2009 | 1187 (604, 2290) | 1154 (588, 2227) | 1124 (572, 2170) | 590 (299, 1151) | 67 (34, 128) | 67 (34, 128) | 66 (34, 128) | 42 (21, 81) | 176 (90, 339) | 171 (87, 329) | 167 (85, 322) | 90 (46, 174) |
| 2010 | 1149 (585, | 1118 (569, | 1089 (554, | 566 (287,110 | 66 (34, 127) | 66 (34, 128) | 66 (34, 127) | 41 (21, 79) | 161 (82, 310) | 156 (80, 301) | 153 (78, 295) | 82 (42, 160) |

| | 2220) | 2162) | 2105) | 5) | | | | | | | | |
|---|---|---|---|---|---|---|---|---|---|---|---|---|
| 2011 | 1151 (586, 2227) | 1121 (570, 2167) | 1089 (554, 2106) | 558 (283, 1091) | 65 (33, 125) | 65 (33, 125) | 64 (33, 124) | 39 (20, 76) | 166 (84, 320) | 161 (82, 310) | 156 (80, 302) | 82 (42, 160) |
| 2012 | 1163 (591, 2249) | 1132 (576, 2190) | 1101 (560, 2131) | 583 (295, 1138) | 68 (35, 131) | 67 (34, 130) | 67 (34, 130) | 42 (21, 81) | 176 (90, 340) | 171 (87, 331) | 168 (86, 324) | 90 (46, 176) |
| 2013 | 1159 (589, 2242) | 1125 (572, 2177) | 1093 (556, 2116) | 588 (297, 1148) | 65 (33, 126) | 66 (33, 126) | 65 (33, 126) | 41 (21, 80) | 176 (90, 340) | 171 (87, 331) | 167 (85, 323) | 89 (45, 174) |
| 2014 | 1153 (586, 2233) | 1120 (569, 2170) | 1088 (553, 2109) | 586 (296, 1145) | 61 (31, 118) | 61 (31, 119) | 61 (31, 118) | 39 (20, 75) | 173 (88, 334) | 168 (85, 325) | 163 (83, 316) | 87 (44, 170) |
| 2015 | 1092 (554, 2118) | 1056 (536, 2049) | 1024 (520, 1987) | 556 (281, 1087) | 61 (31, 119) | 62 (31, 119) | 61 (31, 119) | 39 (20, 75) | 168 (86, 327) | 164 (83, 317) | 160 (81, 310) | 86 (43, 168) |
| 2016 | 1098 (558, 2132) | 1064 (540, 2067) | 1031 (523, 2002) | 562 (284, 1100) | 60 (31, 117) | 60 (31, 117) | 60 (30, 116) | 37 (19, 73) | 177 (90, 343) | 172 (87, 334) | 168 (85, 326) | 90 (46, 176) |
| 2017 | 1074 (545, 2085) | 1039 (528, 2019) | 1006 (510, 1954) | 547 (276, 1071) | 61 (31, 118) | 61 (31, 117) | 60 (31, 117) | 37 (19, 73) | 178 (90, 345) | 172 (88, 334) | 168 (85, 325) | 90 (46, 176) |
| 2018 | 1028 (522, 1999) | 995 (505, 1934) | 962 (488, 1871) | 527 (266, 1032) | 58 (29, 113) | 58 (29, 112) | 58 (29, 112) | 36 (18, 71) | 171 (87, 332) | 166 (84, 322) | 162 (82, 314) | 88 (44, 172) |
| 2019 | 953 (483, 1855) | 919 (466, 1790) | 888 (450, 1729) | 493 (249, 967) | 55 (28, 107) | 55 (28, 106) | 54 (27, 105) | 35 (18, 68) | 168 (85, 328) | 163 (83, 317) | 158 (80, 307) | 86 (43, 169) |
| 2020 | 981 (497, 1913) | 943 (478, 1840) | 907 (459, 1770) | 504 (255, 991) | 61 (31, 119) | 61 (31, 119) | 60 (31, 118) | 39 (20, 76) | 207 (105, 403) | 200 (101, 390) | 194 (98, 379) | 105 (53, 206) |

*Table S7*: Race-specific mortality attributable to PM$_{2.5}$ when conducting our analysis at the blockgroup-level between 2000-2020 using a single CRF, versus a racial/ethnic-group-specific CRF (listed in **Table 1**)

| Year | White | | Black | | Hispanic | |
|---|---|---|---|---|---|---|
| | One CRF | Race-specific CRF | One CRF | Race-specific CRF | One CRF | Race-specific CRF |
| 2000 | 1118 (1089, 1146) | 973 (929, 1002) | 55 (53, 56) | 140 (135, 145) | 126 (123, 129) | 163 (129, 195) |
| 2001 | 1255 (1222, 1287) | 1092 (1043, 1125) | 57 (55, 58) | 144 (139, 149) | 150 (146, 154) | 193 (154, 231) |
| 2002 | 1250 (1218, 1282) | 1088 (1039, 1121) | 56 (55, 58) | 144 (139, 149) | 151 (147, 154 | 194 (154, 233) |
| 2003 | 1213 (1182, 1245) | 1056 (1008, 1088) | 57 (56, 59) | 147 (141, 152) | 154 (150, 158) | 199 (158, 239) |
| 2004 | 1109 (1080, 1137) | 965 (921, 994) | 49 (48, 51) | 127 (122, 131) | 150 (146, 153) | 193 (153, 231) |
| 2005 | 1127 | 980 | 54 | 139 | 153 | 197 |

| Year | | | | | | |
|------|---|---|---|---|---|---|
|      | (1098, 1156) | (936, 1010) | (53, 56) | (134, 144) | (149, 157) | (157, 236) |
| 2006 | 1146 (1117, 1176) | 997 (952, 1027) | 51 (49, 52) | 130 (125, 135) | 152 (148, 156) | 196 (156, 235) |
| 2007 | 1239 (1207, 1271) | 1078 (1029, 1111) | 58 (56, 59) | 147 (142, 152) | 163 (159, 168) | 211 (168, 253) |
| 2008 | 1140 (1111, 1170) | 992 (947, 1022) | 54 (52, 55) | 138 (132, 143) | 161 (157, 165) | 208 (165, 249) |
| 2009 | 1078 (1050, 1106) | 938 (895, 966) | 50 (49, 52) | 130 (125, 135) | 155 (151, 160) | 201 (160, 241) |
| 2010 | 1102 (1073, 1130) | 958 (915, 987) | 54 (53, 56) | 140 (135, 145) | 150 (146, 154) | 194 (154, 232) |
| 2011 | 1084 (1056, 1112) | 942 (900, 971) | 52 (51, 53) | 134 (129, 139) | 153 (149, 157) | 198 (157, 237) |
| 2012 | 1190 (1159, 1221) | 1035 (988, 1066) | 59 (57, 60) | 151 (145, 156) | 175 (171, 180) | 226 (180, 271) |
| 2013 | 1124 (1095, 1153) | 977 (933, 1007) | 55 (54, 56) | 142 (137, 147) | 167 (163, 172) | 216 (172, 259) |
| 2014 | 1176 (1146, 1207) | 1023 (977, 1054) | 57 (55, 58) | 146 (141, 151) | 173 (169, 178) | 224 (178, 268) |
| 2015 | 1133 (1104, 1163) | 985 (941, 1015) | 58 (57, 60) | 151 (145, 156) | 172 (168, 176) | 222 (176, 267) |
| 2016 | 1127 (1097, 1156) | 980 (935, 1009) | 56 (54, 57) | 144 (139, 150) | 178 (173, 183) | 230 (183, 276) |
| 2017 | 1191 (1160, 1222) | 1036 (989, 1067) | 61 (59, 62) | 157 (151, 163) | 193 (188, 198) | 249 (198, 299) |
| 2018 | 1255 (1222, 1183) | 1091 (1042, 1124) | 65 (64, 67) | 168 (162, 175) | 206 (201, 212) | 266 (212, 319) |
| 2019 | 1153 (1123, 1183) | 1002 (957, 1033) | 65 (63, 67) | 168 (162, 175) | 201 (196, 207) | 260 (207, 312) |
| 2020 | 1624 (1582, 1666) | 1413 (1349, 1456) | 99 (96, 101) | 254 (245, 263) | 327 (319, 336) | 423 (336, 507) |

**Table S8**: Race-specific mortality attributable to $NO_2$ when conducting our analysis at the blockgroup-level between 2000-2020 using a single CRF, versus a racial/ethnic-group-specific CRF (listed in **Table 1**)

| Year | White | | Black | | Hispanic | |
|------|-------|-------|-------|-------|----------|----------|
|      | One CRF | Race-specific CRF | One CRF | Race-specific CRF | One CRF | Race-specific CRF |
| 2000 | 2142 (2142, 2466) | 2142 (2142, 2466) | 124 (124, 143) | 242 (242, 257) | 245 (245, 282) | 87 (44, 128) |
| 2005 | 2255 (2255, 2595) | 2255 (2255, 2595) | 129 (129, 148) | 251 (251, 266) | 315 (315, 362) | 111 (57, 165) |
| 2006 | 2099<br>(2099, 2417) | 2099<br>(2099, 2417) | 111<br>(111, 127) | 217<br>(217, 231) | 284<br>(284, 327) | 100<br>(51, 148) |
| 2007 | 1976<br>(1976, 2278) | 1976<br>(1976, 2278) | 108<br>(108, 124) | 212<br>(212, 225) | 269<br>(269, 310) | 95<br>(48, 140) |
| 2008 | 1880<br>(1880, 2167) | 1880<br>(1880, 2167) | 106<br>(106, 122) | 209<br>(209, 223) | 270<br>(270, 310) | 95<br>(48, 140) |
| 2009 | 1780<br>(1780, 2053) | 1780<br>(1780, 2053) | 103<br>(103, 118) | 204<br>(204, 217) | 270<br>(270, 311) | 92<br>(47, 136) |
| 2010 | 1726<br>(1726, 1992) | 1726<br>(1726, 1992) | 102<br>(102, 118) | 203<br>(203, 217) | 263<br>(263, 303) | 84<br>(43, 125) |
| 2011 | 1730<br>(1730, 1997) | 1730<br>(1730, 1997) | 100<br>(100, 115) | 199<br>(199, 212) | 241<br>(241, 278) | 86<br>(44, 128) |
| 2012 | 1748<br>(1748, 2017) | 1748<br>(1748, 2017) | 104<br>(104, 120) | 207<br>(207, 221) | 248<br>(248, 286) | 92<br>(47, 137) |
| 2013 | 1737<br>(1737, 2005) | 1737<br>(1737, 2005) | 101<br>(101, 116) | 202<br>(202, 215) | 264<br>(264, 305) | 92<br>(47, 137) |
| 2014 | 1731<br>(1731, 1999) | 1731<br>(1731, 1999) | 95<br>(95, 109) | 190<br>(190, 203) | 264<br>(264, 305) | 90<br>(46, 134) |
| 2015 | 1633<br>(1633, 1887) | 1633<br>(1633, 1887) | 95<br>(95, 110) | 192<br>(192, 204) | 259<br>(259, 299) | 88<br>(44, 131) |
| 2016 | 1647<br>(1647, 1903) | 1647<br>(1647, 1903) | 93<br>(93, 107) | 187<br>(187, 200) | 253<br>(253, 292) | 92<br>(47, 137) |
| 2017 | 1609<br>(1609, 1859) | 1609<br>(1609, 1859) | 94<br>(94, 108) | 189<br>(189, 201) | 267<br>(267, 308) | 93<br>(47, 137) |
| 2018 | 1540<br>(1540, 1780) | 1540<br>(1540, 1780) | 90<br>(90, 103) | 181<br>(181, 193) | 257<br>(257, 300) | 89<br>(45, 132) |
| 2019 | 1424<br>(1424, 1647) | 1424<br>(1424, 1647) | 84<br>(84, 98) | 172<br>(172, 183) | 252<br>(252, 292) | 87<br>(44, 130) |
| 2020 | 1464<br>(1464, 1693) | 1464<br>(1464, 1693) | 94<br>(94, 109) | 192<br>(192, 206) | 310<br>9310, 359) | 107<br>(54, 159) |